\documentclass[lettersize,journal]{IEEEtran}
\usepackage{amsmath,amsfonts}
\usepackage{algorithmic}
\usepackage{algorithm}
\usepackage{array}
\usepackage[caption=false,font=footnotesize,labelfont=rm,textfont=rm]{subfig}
\usepackage{textcomp}
\usepackage{stfloats}
\usepackage{url}
\usepackage{verbatim}
\usepackage{graphicx}
\usepackage{cite}
\usepackage{mathptmx} 
\usepackage{subfloat}
\usepackage{color}
\usepackage{tikz}

\usepackage{makecell}
\usepackage{multirow}
\usepackage{fancyhdr}
\usepackage{graphicx}
\usepackage{array}
\newcommand{\preserveBackslash}[1]{\let\temp=\\#1\let\\=\temp}
\newcolumntype{C}[1]{>{\preserveBackslash\centering}p{#1}}

\usepackage{fancyhdr}
\usepackage[normalem]{ulem}

\AtBeginDocument{%
  \providecommand\BibTeX{{%
    \normalfont B\kern-0.5em{\scshape i\kern-0.25em b}\kern-0.8em\TeX}}}

\definecolor{yellow}{RGB}{255,217,101}      
\definecolor{blue}{RGB}{102,153,255}      
\definecolor{green}{RGB}{146,208,80}      
\newcommand*\circled[1]{\tikz[baseline=(char.base)]{
            \node[text = white, inner color = black, outer color = black, shape=circle,draw,inner sep=0.3pt] (char) {#1};}}
            
\newcommand*\circledy[1]{\tikz[baseline=(char.base)]{
            \node[text = black, inner color = yellow, outer color = yellow, shape=circle,draw,inner sep=0.3pt] (char) {#1};}}
            
\newcommand*\circledb[1]{\tikz[baseline=(char.base)]{
            \node[text = black, inner color = blue, outer color = blue, 
            shape=circle,draw,inner sep=0.3pt] (char) {#1};}}

\begin{document}

\title{CPSAA: Accelerating Sparse Attention using Crossbar-based Processing-In-Memory Architecture}

\author{Huize~Li,~\IEEEmembership{Member,~IEEE,}
        Hai~Jin,~\IEEEmembership{Fellow,~IEEE,}
        Long~Zheng,~\IEEEmembership{Member,~IEEE,}
        Xiaofei~Liao,~\IEEEmembership{Member,~IEEE,}
        Yu Huang,~\IEEEmembership{Member,~IEEE,}
        Cong Liu,~\IEEEmembership{Student Member,~IEEE,}
        Jiahong Xu,~\IEEEmembership{Student Member,~IEEE,}
        Zhuohui Duan,~\IEEEmembership{Member,~IEEE,}
        Dan Chen,~\IEEEmembership{Student Member,~IEEE,}
        Chuangyi Gui,~\IEEEmembership{Student Member,~IEEE}
\thanks{The authors are with the National Engineering Research Center for Big Data Technology and System, Services Computing Technology and System Lab, Cluster and Grid Computing Lab, School of Computer Science and Technology, Huazhong University of Science and Technology, Wuhan, China (e-mail: {\{huizeli, hjin, longzh, xfliao, yuh, congliu, jhxu, zhduan, cdhust, chygui\}@hust.edu.cn}). Accepted by TCAD in 01/05/2023.}
\thanks{This work is supported by the NSFC (No. 61832006). The correspondence of this paper should be addressed to Yu Huang.}}





\maketitle

\begin{abstract}
The attention-based neural network attracts great interest due to its excellent accuracy enhancement. However, the attention mechanism requires huge computational efforts to process unnecessary calculations, significantly limiting the system's performance. To reduce the unnecessary calculations, researchers propose sparse attention to convert some {\em dense-dense matrices multiplication} (DDMM) operations to {\em sampled dense-dense matrix multiplication} (SDDMM) and {\em sparse matrix multiplication} (SpMM) operations. However, current sparse attention solutions introduce massive off-chip random memory access since the sparse attention matrix is generally unstructured.

We propose CPSAA, a novel crossbar-based {\em processing-in-memory} (PIM)-featured sparse attention accelerator to eliminate off-chip data transmissions. First, we present a novel attention calculation mode to balance the crossbar writing and crossbar processing latency. Second, we design a novel PIM-based sparsity pruning architecture to eliminate the pruning phase’s off-chip data transfers. Finally, we present novel crossbar-based SDDMM and SpMM methods to process unstructured sparse attention matrices by coupling two types of crossbar arrays. Experimental results show that CPSAA has an average of 89.6$\times$, 32.2$\times$, 17.8$\times$, 3.39$\times$, and 3.84$\times$ performance improvement and 755.6$\times$, 55.3$\times$, 21.3$\times$, 5.7$\times$, and 4.9$\times$ energy-saving when compare with GPU, FPGA, SANGER, ReBERT, and ReTransformer.
\end{abstract}


\section{Introduction}
\label{intro}
Attention-based neural network shows accuracy leaps in machine learning applications, e. g., {\em natural language processing} (NLP)~\cite{Kalyan21} and computer vision~\cite{Han20}. Different from the commonly used {\em Convolutional Neural Network} (CNN) or {\em Recurrent Neural Network} (RNN) models, Transformer~\cite{Vaswani17} adopts a pure attention-based neural network to better identify the dependencies between tokens of the input sequence. Following this design, Transformer and its variants achieve great accuracy improvement in NLP tasks~\cite{Kalyan21}, such as machine translation~\cite{Vaswani17} and question answering~\cite{Geigle21}, etc. Attention is also widely used in computer vision tasks~\cite{Han20} including image classification~\cite{Chen21} and object detection~\cite{Liu21}, etc. 

The vanilla attention mechanism~\cite{Vaswani17} is usually implemented as DDMM and softmax operations. By computing an attention score matrix, the attention mechanism can pay attention to these relevant token pairs. There is overwhelming computation pressure in processing these irrelevant token pairs, leading to intolerable execution time~\cite{Ham20}. Researchers propose sparse attention by adding a sparsity pruning phase before the attention calculation to reduce irrelevant calculations~\cite{Ham20, Lu21}, since most tokens in the input sequence are unrelated to the current query. There are two types of sparse attention designs, i.e., software-based and software-hardware co-design methods~\cite{Lu21}. Software-based methods~\cite{Zaheer20, Tay20} aim to propose various optimization algorithms to reduce computational overhead by increasing sparsity. Software-hardware co-design solutions accelerate sparse attention by taking advantage of high-parallelism hardware, such as {\em Field Programmable Gate Array} (FPGA)~\cite{Zhang21, Li20} and {\em Application Specific Integrated Circuit} (ASIC)~\cite{Lu21, Ham20, Wang21}.

The above solutions only achieve limited speedups since both the sparsity pruning and attention calculation phases involve many off-chip data transfers. Emerging crossbar-based architectures are promising to solve the off-chip data transmission problem, such as {\em Resistive Random Access Memory} (ReRAM) and {\em ReRAM-based content addressable memory} (ReCAM)~\cite{Kaplan17}. ReCAM is suitable for high parallel comparison, the core operation of content-based similarity search in the attention mechanism. ReRAM is ideal for {\em vector-matrix multiplication} (VMM) operation, which has superior performance handling the DDMM operations of attention-based neural network. Utilizing the in-situ processing ability of ReRAM arrays, there emerge ReRAM-based PIM-featured solutions to accelerate traditional neural network~\cite{Shafiee16} and the dense attention mechanism~\cite{Yang20, Kang21}. However, these PIM-based solutions can hardly extend to accelerate the sparse attention for the following reasons.

{\bf First}, the ReRAM array's write overhead cannot be ignored as it is in ReRAM-based CNN and RNN accelerators. Solving the ReRAM write overhead is urgent because many matrices in the attention mechanism cannot be reused and need to be written in runtime. {\bf Second}, the sparse attention involves the sparsity pruning phase, which is not considered by current dense attention accelerators. Using the current software-based pruning algorithm can promote the PIM-based attention accelerator. However, the software-based pruning methods have poor performance because they need to load all input matrices from the off-chip memory to the processor. Moreover, the sparsity of the attention mechanism is pretty unstructured, which will introduce lots of off-chip random memory access to the attention calculation phase. {\bf Finally}, the sparse attention involves the SDDMM and SpMM operations. Direct application of current ReRAM-based sparse methods~\cite{Lin19, Song18} to ReRAM-based SDDMM and SpMM operations will achieve inferior performance (for details to see $\S~\ref{sdd}$ and $\S~\ref{SpMM}$).
Given this landscape, we propose CPSAA, a novel {\underline C}rossbar-based {\underline P}IM-featured {\underline S}parse {\underline A}ttention {\underline A}ccelerator. First, we design the attention calculation mode to increase the parallelism of CPSAA dataflow while hiding ReRAM writing overhead. Second, we design a novel PIM-based sparsity pruning architecture to support our calculation mode while removing the off-chip data transmissions. Finally, we couple ReRAM and ReCAM arrays and propose new PIM-based SDDMM and SpMM methods to utilize the unstructured sparsity. The main contributions of this paper are as follows:
\begin{itemize}
    \item We propose CPSAA, a PIM-based sparse attention accelerator, to speed up both the pruning and attention calculation phases of the neural network inference. 
    \item We design the pruning phase as a novel PIM architecture to eliminate off-chip data transfers. The attention calculation architecture involves novel ReRAM-based SDDMM and SpMM methods to {increase the parallelism of sparse attention}.
    \item We evaluate and compare CPSAA with state-of-the-art attention mechanism solutions. The experimental results show that CPSAA has performance improvement in speedups and energy-saving.
\end{itemize}


\section{Background and Motivation}
\label{backg}

\begin{figure}[t]
\centering
\includegraphics[width=8.5cm]{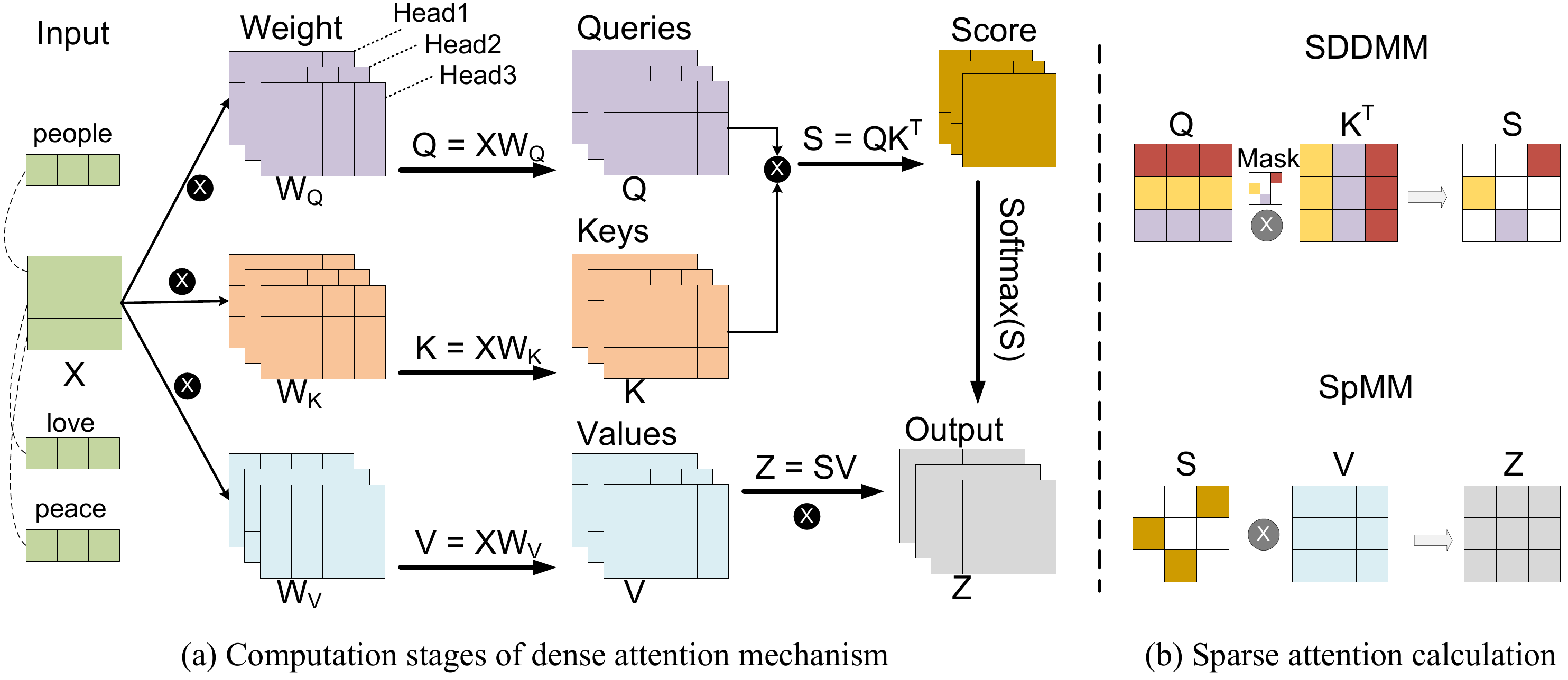}
\caption{Dataflow of attention mechanism}
\label{attention}
\vspace{-1em}
\end{figure}

\subsection{Sparse Attention}
The vanilla attention mechanism maps a query matrix $Q$ and a key-value matrix pair $K$-$V$ to an output matrix $Z$. Figure~\ref{attention} (a) depicts the calculation procedure of the dense attention mechanism. First, the {\em query} ($Q$), {\em key} ($K$), and {\em value} ($V$) matrices are obtained by multiplying the embedded input sequence $X$ with the corresponding weight matrices, W$_Q$, W$_K$, and W$_V$. Next, the score matrix $S$ is calculated by multiplying $Q$ with $K^\mathsf{T}$. Then, the $S$ matrix is normalized with a row-wise softmax function {\bf Softmax($S$)}. Finally, $S$ is multiplied by the $V$ matrix to get the output $Z$.

As Figure~\ref{attention} (a) shows, $S$ reveals the relevance between the tokens of $Q$ and $K$ matrices. Researchers find that most tokens in $Q$ are irrelevant to $K$, making the $S$ matrix inherently sparse~\cite{Ham20}. The above fact makes it possible to utilize the sparsity of $S$ to avoid computing these irrelevant token pairs~\cite{Ham20}. Thus, people propose sparse attention, which adopts a sparsity pruning phase to save computational resources of the attention calculation phase.

\vspace{-1em}
\begin{equation}
G[i,j] = \boldsymbol{binarize}(\Tilde{S}[i,j], \theta) = \left\{
\begin{array}{rcl}
    1,  & {if~\Tilde{S}[i,j] \ge \theta }\\
    0,  & {otherwise}
\end{array}
\right.
\label{exp3}
\end{equation}
\vspace{-1em}

SANGER~\cite{Lu21} proposes a prediction-based pruning method and achieves high sparsity. The idea of their pruning method is to calculate the approximate score matrix $\Tilde{S}$ using a low-precision DDMM operation, $\Tilde{S} = Softmax(\mathbf{QU}^{-1}(\mathbf{QU}(Q)\cdot \mathbf{QU}(K^\mathsf{T}))/\sqrt{d})$, where $d$ is the dimension of the input embeddings, and $\mathbf{QU}(x) = {round}(\gamma x)$ is a quantization operator that maps input buffers to low-bit by a scaling factor $\gamma$ and a rounding bit-shift. $\mathbf{QU}^{-1}(\cdot)$ is the corresponding de-quant operator that transforms low-bit outputs back into high precision. Then, a binarization procedure described in equation~\eqref{exp3} will convert the $\Tilde{S}$ matrix to a binary mask matrix $G$, where $\theta$ is the threshold of this binarization function. Because the sparsity of the mask matrix is similar to the score matrix $S$, SANGER can convert the DDMM operation $S = Q\cdot K^\mathsf{T}$ to SDDMM operation. As Figure~\ref{attention} (b) shows, SDDMM operation generates the sparse score matrix $S$ with three input matrices (two dense matrices $Q$, $K$, and one sparse mask matrix), which can utilize the mask matrix to avoid calculating the zero-value (white cells) of the $S$ matrix. Utilizing the sparse $S$ matrix, people can convert the DDMM operation $Z = S\cdot V$ to the SpMM operation. Figure~\ref{attention} (b) shows the visualization of the SpMM operation, which is a VMM operation between a sparse matrix $S$ and a dense matrix $V$.

\begin{figure}[t]
\centering
\includegraphics[width=8.9cm]{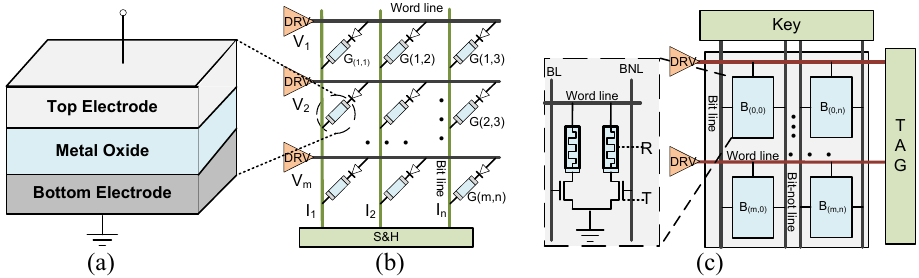}
\vspace{-1em}
\caption{(a) One ReRAM cell, (b) The ReRAM crossbar architecture, and (c) The ReCAM array architecture}
\vspace{-1em}
\label{figure:basic}
\end{figure}

\subsection{ReRAM and ReCAM Basics}
\label{ReRAM}

Figure~\ref{figure:basic} (a) shows one ReRAM cell, which is comprised of a top electrode, a metal oxide layer, and a bottom electrode. Following the Kirchhoff's Current Law, ReRAM crossbar array can process VMM operation efficiently\cite{Wen19}, denoted as $I_n$ = $\sum_{m=0}^{N} V_m \times G_{(m,n)}$, as Figure~\ref{figure:basic} (b) shows. Utilizing this high parallel in-situ VMM operation, several ReRAM-based neural network accelerators~\cite{Shafiee16} and graph processing accelerators~\cite{Song18, Zheng20, Qian20} have been proposed.

The ReCAM array architecture is shown in Figure~\ref{figure:basic} (c), which is comprised of lots of {\em two transistors and two memristors} (2T2R) ReCAM bit-cells~\cite{Kaplan17}. One ReCAM bit-cell contains a couple of ReRAM cells. One ReCAM array contains a Key register, a ReCAM cells array, {\em drivers} (DRV), and {\em tag registers} (TAG). ReCAM arrays can perform vector-scalar comparisons in parallel~\cite{Kaplan17}. The TAG will latch the `1' signal if one row matches with the Key register.

\begin{table*}[t]
\begin{center}
    \caption{Current attention mechanism solutions}
    \label{tab:existing}
    \small
    \vspace{-0.5em}
    \begin{tabular}{|p{2cm}<{\centering}|p{1.7cm}<{\centering}|p{1.9cm}<{\centering}|p{1.70cm}<{\centering}|p{1.70cm}<{\centering}|p{1.7cm}<{\centering}|p{1.70cm}<{\centering}|p{1.7cm}<{\centering}|}
    \hline
    \multirow{2}*{\textbf{Features}} &\multicolumn{2}{c|}{\textbf{Hardware-based}}&  \multicolumn{3}{c|}{\textbf{Software-hardware co-design}} & \multicolumn{2}{c|}{\textbf{Software-based}} \\
    \cline{2-8}
    {}&{\cite{Yang20, Kang21}} & {CPSAA} & \cite{Li20, Zhang21} & \cite{Ham20, Wang21}&{\cite{Lu21, Qu22}}&{\cite{Child19, Zaheer20}}&{\cite{Cui19, Zhao19}}\\
    \cline{1-8}
    {\textbf{Spar. pattern}} & {dense} & {dynamic} & {static} & {dynamic} & {dynamic} & {static} & {dynamic}\\
    \cline{1-8}
    {\textbf{Pruning plt.}} & {-} & {ReRAM} & {CPU/GPU} & {CPU/GPU} & {CPU/GPU}& {CPU/GPU} & {CPU/GPU}\\
    \cline{1-8}
    {\textbf{Attention plt.}} & {ReRAM} & {ReRAM} & {FPGA} & {ASIC} & {ASIC}& {CPU/GPU} & {CPU/GPU}\\
    \cline{1-8}
    {\textbf{Sparsity visualization}} & {$\includegraphics[width = 1.6cm]{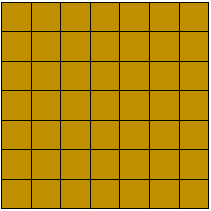}$}&{$\includegraphics[width = 1.6cm]{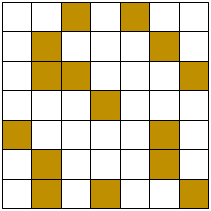}$}&{$\includegraphics[width = 1.6cm]{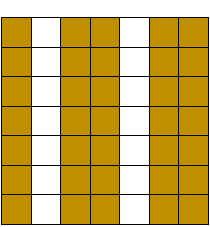}$}&{$\includegraphics[width = 1.6cm]{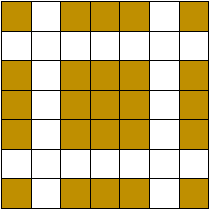}$}&{$\includegraphics[width = 1.6cm]{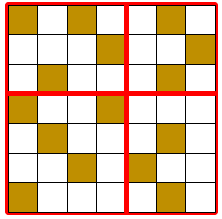}$}&{$\includegraphics[width = 1.6cm]{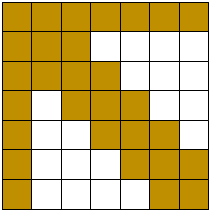}$}&{$\includegraphics[width = 1.6cm]{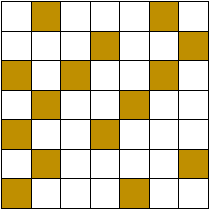}$}\\
    \cline{1-8}
    {\textbf{Sparsity regularity}} & {-}&{unstructured}&{coarse-grain structured}&{coarse-grain structured}&{fine-grain structured}&{coarse-grain structured}&{unstructured}\\
    \cline{1-8}
    {\textbf{Sparsity}} & {-}&{high}&{low}&{medium}&{high}&{low}&{high}\\
    \cline{1-8}
    {\textbf{Speedup}} & {high}&{high}&{low}&{medium}&{high}&{low}&{low}\\
    \hline
    \end{tabular}
\end{center}
\vspace{-2em}
\end{table*}

\subsection{Related Work}
\label{exist}
Table~\ref{tab:existing} lists the recent studies of the attention mechanism. Different from SANGER's classification, we divide these works into three categories, i.e., software-based, software-hardware co-design, and the new adding hardware-based. The software-based methods can be further divided into static and dynamic sparsity patterns. The static sparsity pattern can only get coarse-grained sparsity for data dependent~\cite{Child19, Zaheer20}. The dynamic sparsity pattern can achieve higher sparsity conditioned on individual input samples during runtime, but their unstructured sparsity increases the random memory access overhead~\cite{Cui19, Zhao19}.



Software-hardware co-design methods adopt more efficient hardware, such as FPGAs and ASICs. FTRANS~\cite{Li20} is an efficient acceleration framework for transformer-based NLP. Zhang et al. propose a novel pruning method and design the FPGA-based sparse attention accelerator~\cite{Zhang21}. A$^3$~\cite{Ham20} can accelerate the attention mechanism with algorithmic approximation and hardware specialization. SpAtten~\cite{Wang21} leverages token sparsity, head sparsity, and quantization opportunities to reduce redundent computation and random access. SANGER~\cite{Lu21} presents a prediction-based pruning algorithm to reduce unnecessary calculations while designing the ``splitting and packing" algorithm to reduce the massive random memory access overhead. {DOTA~\cite{Qu22} proposes a dynamic weak connections detector to avoid unnecessary attention calculation and promote sparse attention.} 

Although these co-design solutions can relieve the random memory access by designing hardware-specific algorithms, many off-chip data transmissions exist because of their separate memory and processor architecture. Using SANGER as an example, the ``splitting and packing" algorithm can reduce random memory access by converting the unstructured sparsity to fine-grained structured sparsity. {However, the benefit of the fine-grained structured sparsity comes at the cost of the dynamically configured control signals, which are highly complex to scheduling processing elements and memory access in runtime. In addition, SANGER introduces off-chip random access and data dependency to their sparsity pruning phase.} Thus, no current accelerators can solve the off-chip random memory access problem elegantly.

We also list current hardware-based PIM-featured dense attention accelerators in Table~\ref{tab:existing}. These solutions use high parallel ReRAM arrays to significantly reduce the latency of DDMM operations~\cite{Yang20, Kang21}. However, since the sparsity of the attention matrices is not considered, current PIM-based attention accelerators need to calculate all five DDMM operation steps, wasting lots of computational resources on irrelevant tokens. Therefore, how to design a PIM-based sparse attention accelerator and reducing unnecessary calculations become the key to achieve further accelerations.

\begin{figure}[t]
\centering
\includegraphics[width=8.7cm]{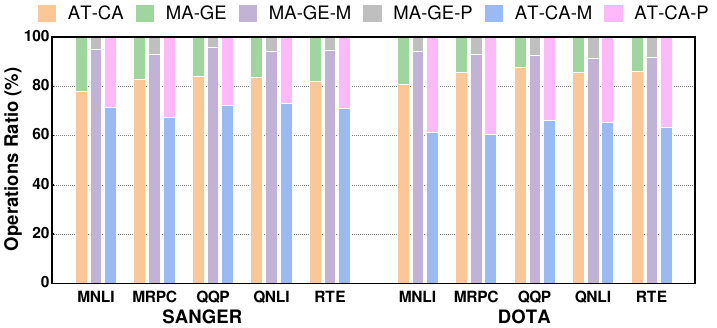}
\caption{Response time breakdown of SANGER and DOTA}
\label{sanger_break}
\vspace{-1em}
\end{figure}


\subsection{Motivation}
\label{motivation}
{\em We find that current sparse attention accelerators have many off-chip data transmissions.} We design experiments on two sparse attention accelerators, i.e., SANGER and DOTA, to confirm this statement. Figure~\ref{sanger_break} presents the operations ratio of SANGER \cite{Lu21} and {DOTA~\cite{Qu22}} running five real-world datasets (details to see Section V). The ratio of the operation comes from breaking down the response time to the {\em mask matrix generation} (MA-GE) and the {\em attention mechanism calculation} (AT-CA). To assess the impact of off-chip memory access, we further break down the MA-GE and AT-CA to the processor execution time (MA-GE-P and AT-CA-P) and the memory access time (MA-GE-M and AT-CA-M). The memory access time is obtained by subtracting the average kernel runtime from the total execution time.

{\bf First, the overhead of the mask generation and attention calculation phases both cannot be ignored.} {Figure~\ref{sanger_break} shows the MA-GE takes an average of 17.9\% (14.3\% in DOTA) response time, while the AT-CA takes 82.1\% (85.7\% in DOTA).} The calculation overhead of the pruning phase is smaller than the attention calculation since the pruning phase of SANGER and DOTA uses low-precision computing.

{\bf Second, most of the overhead of the mask generation phase comes from the off-chip memory access.} The MA-GE-M takes an average of 94.6\% {(92.7\% in DOTA)} response time while the MA-GE-P takes 5.4\% {(7.3\% in DOTA).} The reason is that the MA-GE of SANGER (DOTA) needs to access $Q$ and $K$ from the off-chip memory, which involves many off-chip data transfers and takes lots of response time. 

{\bf Third, the attention calculation also spends a lot of time on off-chip data transfers.} The AT-CA-M takes an average of 71.2\% {(63.5\% in DOTA)} response time while the AT-CA-P takes 28.8\% {(36.5\% in DOTA).} That is because all the input matrices, $Q$, $K$, $V$, and $S$, need to be sent from the off-chip memory to the ASIC-based processor. These off-chip data transmissions greatly increase access latency and further hurts performance. The massive off-chip data transmissions motivate us to design a novel sparse attention accelerator to simultaneously accelerate MA-GE and AT-CA with pretty low off-chip random access.

\section{Calculation Mode}
\label{mode}
\begin{figure}[t]
\centering
\includegraphics[width=8.7cm]{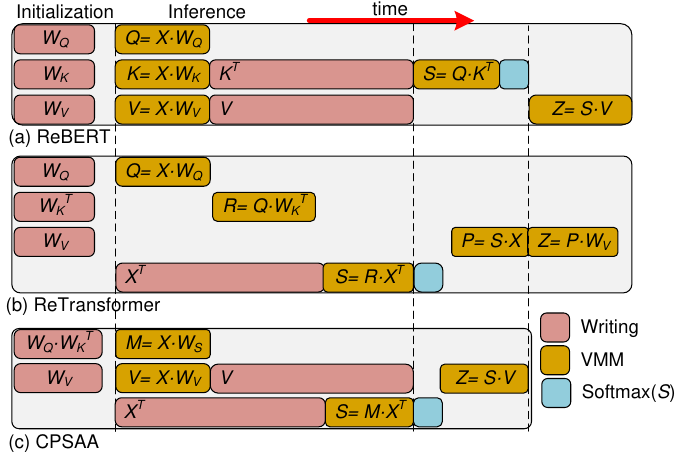}
\caption{The calculation mode of (a) ReBERT, (b) ReTransformer, and (c) dense-version CPSAA}
\label{compare}
\vspace{-1em}
\end{figure}

In conventional ReRAM-based CNN and RNN accelerators, the write overhead of their weight matrices can be ignored because these matrices can be reused for different inputs. Unfortunately, the pretty high ReRAM write overhead of the $K$ and $V$ matrices in the attention mechanism cannot be ignored because these matrices vary for different input sequences. Even worse, the write operation works serial with the VMM operation and greatly increases the latency, just as the calculation mode of ReBERT~\cite{Kang21} shown in Figure~\ref{compare} (a). ReBERT uses the same calculation mode as SANGER, which has pretty high VMM parallelism since it can get $Q$, $K$, and $V$ simultaneously. However, the VMM operation $S= Q\cdot K^\mathsf{T}$ has to wait for the write operations of matrices $K^\mathsf{T}$, which greatly hurts the performance of ReBERT.

To reduce the performance impact of ReRAM writing, ReTransformer~\cite{Yang20} depicts a novel calculation mode as Figure~\ref{compare} (b) shows. To avoid the VMM operations waiting a long time for the writing of $K^\mathsf{T}$, the authors avoid generating $K$ and $V$. With the weight matrices $W_Q$, $W_K$, and $W_V$ used in serial, their new computational mode achieves the goal that the VMM operations can work concurrently with the writing. {However, three challenges limit this calculation mode to efficiently support sparse attention: 1) \textit{Data dependency.} As Figure~\ref{compare} (b) shows, ReTransformer has to sequentially get matrix $Q \rightarrow R \rightarrow S \rightarrow P \rightarrow Z$ with high data dependency, while $Q$, $K$, and $V$ matrices could be parallel computed in ReBERT. 2) \textit{Poor performance when extending to mask generation.} Things get worse in sparse attention because there is a mask generation phase. Suppose ReTransformer wants to get the quantized $S$ (i.e., mask matrix). ReTransformer should get $Q$ first; then use the quantized $Q$ to get the quantized $R$; finally, to get quantized $S$. Therefore, the generation of $Q$ (DDMM operation) becomes the critical path of the mask generation. 3) \textit{Data dependency between the mask generation and attention calculation.} The mask generation of ReTransformer must wait for its attention calculation (i.e., data dependency). Because ReTransformer must generate $Q$ and $R$ matrices to get the mask matrix. Therefore, the mask generation of ReTransformer can not work concurrently with its attention calculation, i.e., ReTransformer does not see the potential of parallel execution between mask generation and attention calculation.}

\vspace{-1em}
\begin{equation}
S = Q\cdot K^\mathsf{T} = (X\cdot W_Q)\cdot(X\cdot W_K)^\mathsf{T}
\label{exp1}
\end{equation}
\vspace{-1em}

\vspace{-1em}
\begin{equation}
S = (X\cdot W_Q)\cdot(W_K^\mathsf{T}\cdot X^\mathsf{T}) = X\cdot (W_Q\cdot W_K^\mathsf{T})\cdot X^\mathsf{T}
\label{exp2}
\end{equation}

{We develop a novel calculation mode to solve the above challenges, as Figure~\ref{compare} (c) shows. We find that the above three challenges are the victims of the data dependency of $Q$. To unbind the data dependency of $Q$, we pre-calculate the weight matrix $W_S$ by performing $W_S = W_Q\cdot W_K^\mathsf{T}$, and we pre-store $W_S$ and $W_V$ in ReRAM arrays. The calculation of $S$ described in equation~\eqref{exp1} can be converted to equation~\eqref{exp2} due to the combination law of matrix multiplication. Based on equation~\eqref{exp2}, we can get the score matrix $S$ by performing two VMM operations, i.e., $M = X\cdot W_S$ and $S = M\cdot X^\mathsf{T}$. To further unbind the data dependency between $P$ and $Z$ in Figure~\ref{compare} (b), we calculate $V$ in advance as Figure~\ref{compare} (c) shows. After $V = X\cdot W_V$ is calculated, we can get the output $Z$ matrix by performing $S\cdot V$. Our novel computational mode can unbind the data dependency of $Q$ and achieve parallel VMM operations, while the VMM operations can also run parallel with write operations in an asynchronous manner. Further, this new calculation mode can also significantly improve the performance of mask generation (for details to see Figure~\ref{ourdataflow}).}

\begin{figure}[t]
\centering
\includegraphics[width=8.9cm]{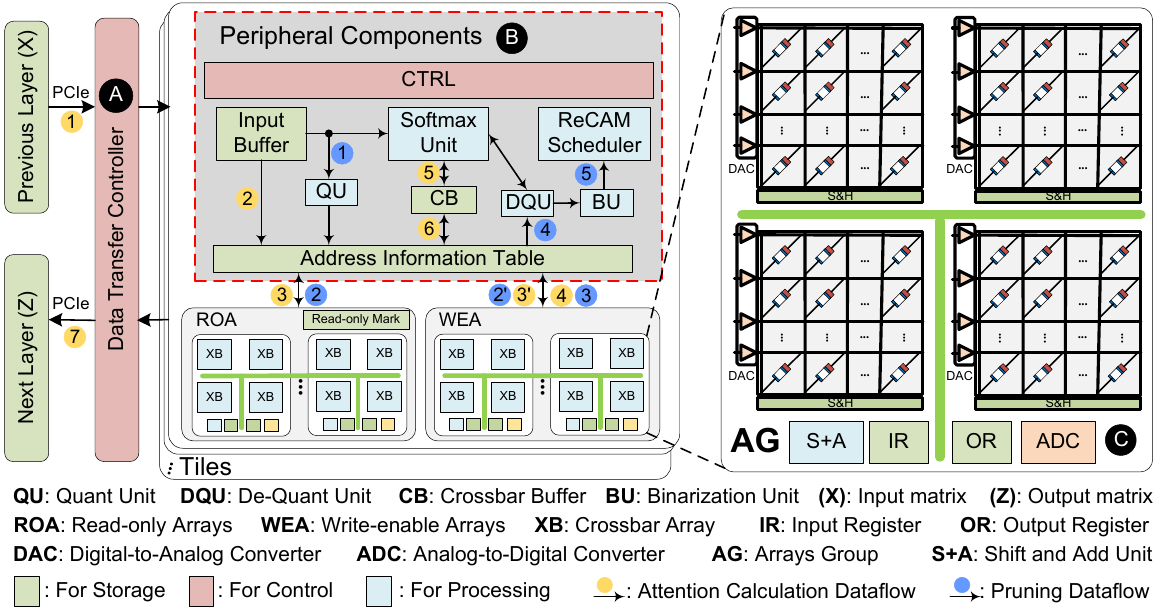}
\caption{CPSAA architecture}
\label{architecture}
\end{figure}

\begin{figure}[t]
\centering
\includegraphics[width=8.5cm]{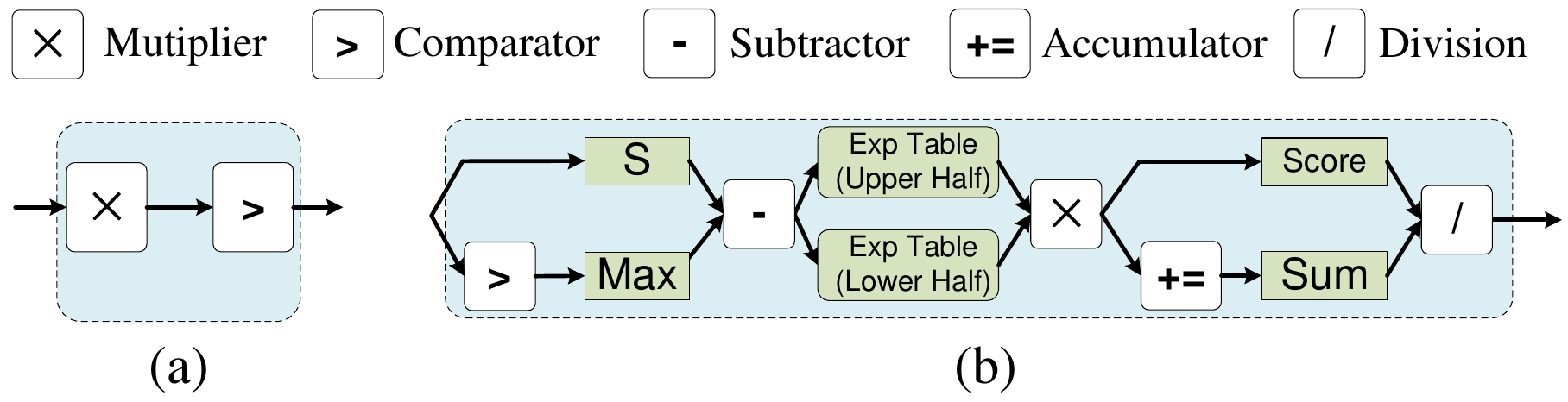}
\caption{(a) Details of QU, (b) details of SU}
\label{SU}
\vspace{-1em}
\end{figure}

\section{CPSAA}
\label{archi}
\subsection{Overview}
\begin{figure*}[t]
\centering
\includegraphics[width=14cm]{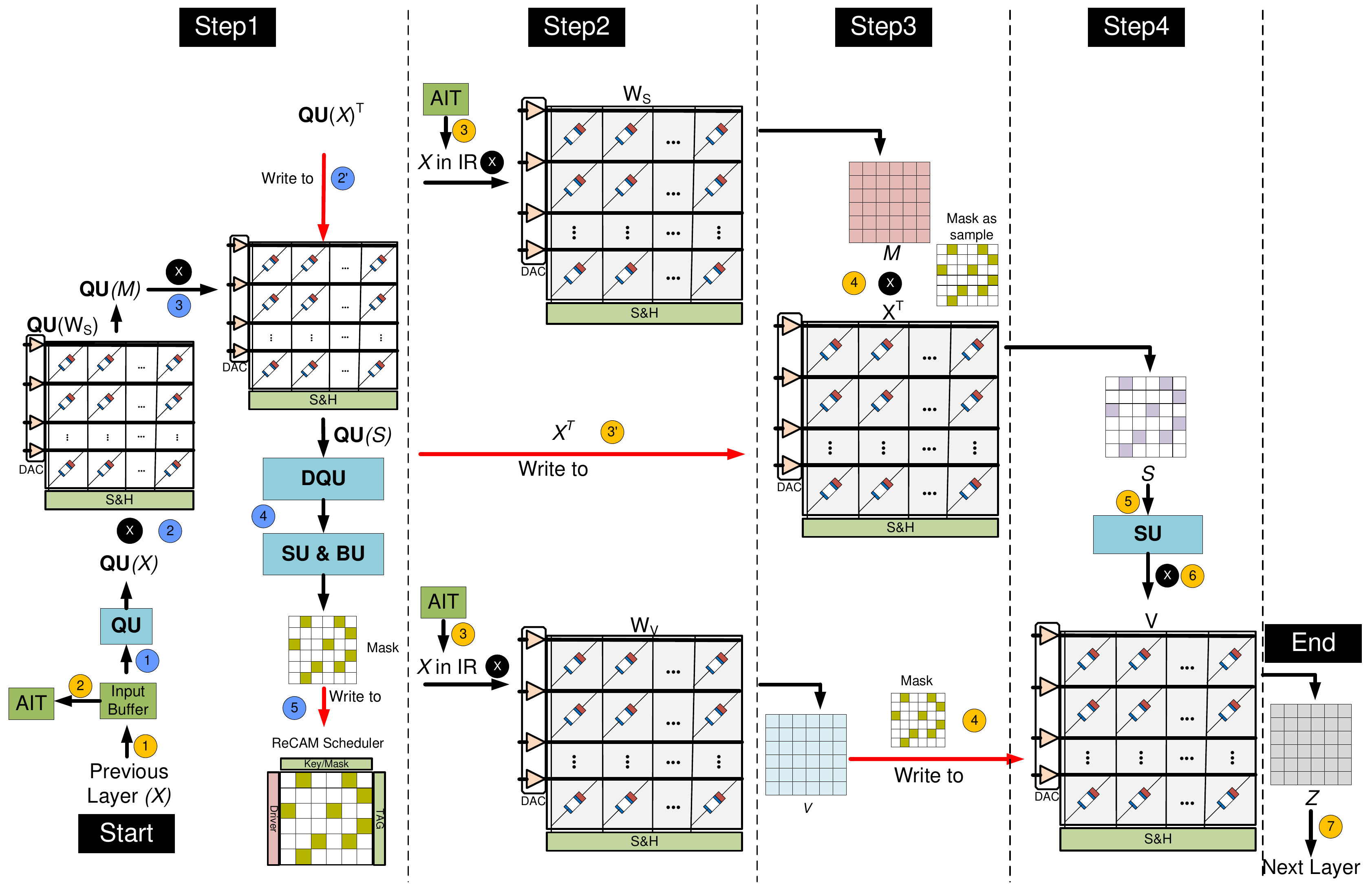}
\caption{CPSAA visualization dataflow}
\label{ourdataflow}
\vspace{-1em}
\end{figure*}

Figure~\ref{architecture} shows the overview architecture of CPSAA, which contains several Tiles, and each Tile has two main parts, i.e., the peripheral components (red-dotted rectangle \circled{B}) and ReRAM arrays. The {\em peripheral components} (PC) contain a {\em controller} (CTRL), several Buffers, {\em Quant and De-Quant Units} (QU and DQU), one {\em Softmax Unit} (SU), one {\em Binarization Unit} (BU), and {two ReCAM arrays worked as Scheduler}. The ReRAM arrays are classified as {\em read-only arrays} (ROA) and {\em write-enable arrays} (WEA). The parameters obtained from pre-training are evenly distributed to the read-only arrays in each tile of CPSAA. The tiles assigned to the same attention head communicate with each other through on-chip bus. Because real-world applications usually need multiple attention layers working together, we also show the input matrix from the previous layer and the output matrix to the next layer, which is the off-chip data transmission managed by the {\em data transfer controller} (DTC~\circled{A}). We use the circled color numbers to show the CPSAA dataflow, which is corresponded to the numbers in Figure~\ref{ourdataflow}. The functions of all components are as follows:

{\bf CTRL.} The controller will generate control signals for all components in one Tile.

{\bf Buffers.} The {\em Input Buffer} (IB) will store the input embeddings, i.e., $X$ matrices. The {\em Crossbar Buffer} (CB) will store the matrices generated at running time. The {\em address information table} (AIT) will record the address information of matrices in the ReRAM arrays, which will be useful when we need to know the location of a specific sub-matrix.

{\bf QU and DQU.} The Quant Unit will quantize the matrix to a low-precision representation (via $\mathbf{QU}(\cdot)$ function), and the De-quant Unit will convert the low-precision representation back to high precision (via $\mathbf{QU}^{-1}(\cdot)$ function). The architecture of the QU is shown in Figure~\ref{SU} (a), and the DQU is designed in the inverse process.

{\bf SU and BU.} The Softmax Unit will perform the softmax function, and the detailed architecture of the SU is the same as A$^3$~\cite{Ham20}, shown in Figure~\ref{SU} (b). The input of SU is matrix $S$, which is calculated by multiple crossbar arrays in one tile. While crossbar arrays calculate matrix $S$ in different tiles, CPSAA will gather these results to generate $S$. The Binarization Unit will convert a given matrix to the `01' matrix, which is performed by a binarization comparator.

{\bf ReCAM {Scheduler}.} The ReCAM arrays will store the mask matrices, which will be used to generate control signals for the ReRAM-based VMM operations.

{\bf ROA.} The read-only arrays will pre-store those matrices that can be reused for different inputs, i.e., $\mathbf{QU}(W_S)$, $W_S$, and $W_V$. The ROA contains several {\em Arrays Group} (AG), and we set a read-only mark for the ROA.

{\bf WEA.} The write-enable arrays will store these matrices that are generated in the runtime, i.e., $\mathbf{QU}(X^\mathsf{T})$, $X^\mathsf{T}$, and $V$.

{\bf AG.} The Arrays Group (\circled{C}) is the basic storage and computing units, which contains one {\em shift and add Unit} (S+A), one {\em input register} (IR), one {\em output register} (OR), one {\em analog-digital-converter} (ADC), and several ReRAM arrays.

\subsection{Dataflow}
\label{sparse_flow}
The color-circled numbers in Figure~\ref{ourdataflow} show two CPSAA visualization dataflows, i.e., the pruning phase's dataflow (blue-circled numbers) and attention calculation dataflow (yellow-circled numbers). To illustrate the dataflow more clearly, we divide the CPSAA dataflows into four steps, as Figure~\ref{ourdataflow} shows. Step1 is a PIM-based pruning method to generate the mask matrix. Step2 contains the calculation of $M= X\cdot W_S$, $V= X\cdot W_V$, and the write of the matrix $X^\mathsf{T}$ ({Noting that Step1 and Step2 can run in parallel since we remove the datapath dependency of $Q$ between these two steps}). Step3 contains the SDDMM operation $S= M\cdot X^\mathsf{T}$ and the write of matrix $V$. Step4 performs the SpMM operation $Z= S\cdot V$. It is worth noting that Step2-Step4 is the detailed version of the computational mode in Figure~\ref{compare} (c).


{\bf Pruning Dataflow.} The pruning algorithm of SANGER needs to use the full-precision $Q$ and $K$ matrices to generate the mask matrix, which makes the pruning phase wait for the calculation of these intermediate matrices $Q$ and $K$. To avoid the waiting of the pruning phase ({i.e., Step1 waiting for Step2}), we fine-tune the mask generating algorithm as equation~\eqref{exp4}, where $Bi(\cdot)$ is the binarization function and $Soft(\cdot)$ is the softmax function, and the ${\mathbf{QU}}(X)$, ${\mathbf{QU}}(W_S)$, and ${\mathbf{QU}}(X^\mathsf{T})$ denotes the low-precision $X$, $W_S$, and $X^\mathsf{T}$. Our new pruning algorithm can use the input embedding matrix and the weight matrix to calculate the mask matrix, {which does not need to use the results of Step2, i.e., $M$ or $V$}.




\vspace{-1em}
\begin{equation}
mask = \boldsymbol{Bi}(\boldsymbol{Soft}({\mathbf{QU}^{-1}}({\mathbf{QU}}(X)\mathbf{QU}(W_S)\mathbf{QU}(X^\mathsf{T})) / \sqrt{d}))
\label{exp4}
\end{equation}
\vspace{-1em}

As Step1 in Figure~\ref{ourdataflow} shows, we pre-store the low-precision $\mathbf{QU}(W_S)$ in ROA. When an input matrix $X$ arrives at CPSAA, the system will send $X$ to the QU and get ${\mathbf{QU}}(X)$ (\circledb{1}). Then, ${\mathbf{QU}}(X^\mathsf{T})$ will be written to WEA (\circledb{2'}), and a VMM operation will be performed between ${\mathbf{QU}}(X)$ and ${\mathbf{QU}}(W_S)$ to get ${\mathbf{QU}}(M)$ (\circledb{2}). Next, the system will perform a VMM operation between ${\mathbf{QU}}(M)$ and ${\mathbf{QU}}(X^\mathsf{T})$ to generate ${\mathbf{QU}}(S)$ (\circledb{3}). After that, ${\mathbf{QU}}(S)$ will be sent to the DQU, SU, and BU to get the mask matrix (\circledb{4}). Finally, the mask matrix will be written to a ReCAM array (\circledb{5}). {The sparsity of the mask is similar to the matrix $S$}, so we can use the mask matrix to convert the DDMM operation $S = M\cdot X^\mathsf{T}$ to the SDDMM operation.





{\bf Attention Calculation Dataflow.} As shown in Figure~\ref{ourdataflow}, the weight matrices $W_S$ and $W_V$ are pre-stored in ROA. At the beginning, CPSAA receives and stores an input embedding matrix $X$ to the Input Buffer (\circledy{1}). Then, the following three phases will be performed in parallel: 1) CPSAA will write $X^\mathsf{T}$ to WEA (\circledy{3'}), 2) CPSAA will send $X$ to the IR via AIT to {perform the DDMM operation} $M = X\cdot W_S$ (\circledy{2}\circledy{3}), and 3) CPSAA will perform the DDMM operation $V = X\cdot W_V$ to generate matrix $V$ (\circledy{2}\circledy{3}). After that, CPSAA can use mask matrix to perform the SDDMM operation $S = M\cdot X^\mathsf{T}$ to get the score matrix $S$ while writing $V$ to WEA (\circledy{4}). After CPSAA performs the softmax function on $S$ (\circledy{5}), CPSAA will use the mask matrix to perform a SpMM operation $Z = S\cdot V$ to get the final result (\circledy{6}\circledy{7}).

\begin{figure}[t]
\centering
\includegraphics[width=8.9cm]{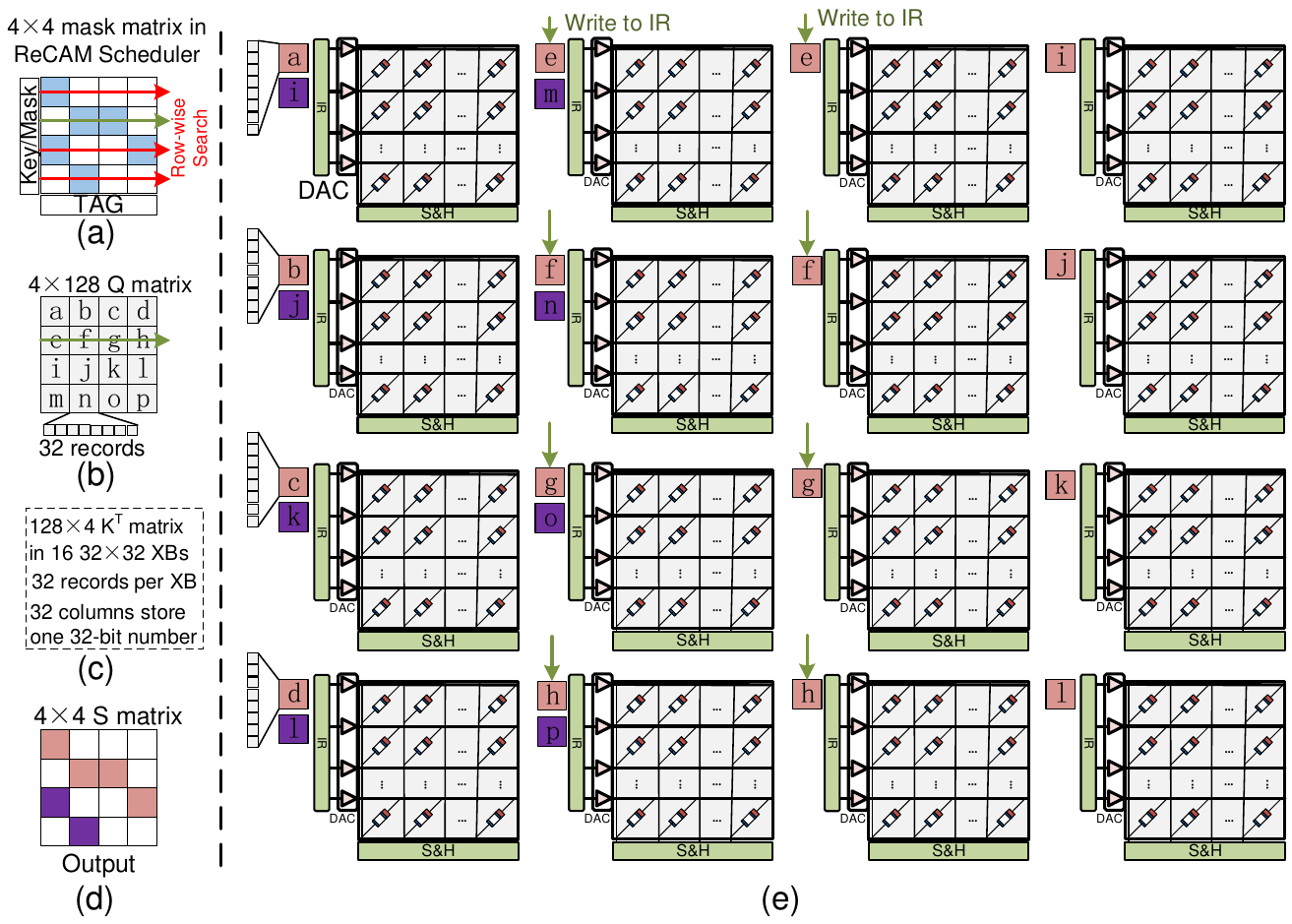}
\caption{Novel ReRAM-based SDDMM method with (a) the mask matrix in the ReCAM Scheduler, (b) the 4$\times$128 $Q$ matrix, (c) mapping rules of $K^\mathsf{T}$ matrix, which is a brief introduction to the $K^\mathsf{T}$ matrix in Figure~\ref{QKmeth} (e), (d) parallel-accessible elements in $S$ matrix, and (e) visualization of the SDDMM operation in CPSAA}
\label{QKmeth}
\vspace{-1em}
\end{figure}

\subsection{SDDMM Method}
\label{sdd}
CPSAA can use the mask matrix to convert the DDMM operation to SDDMM operation. {However, current SDDMM solutions such as DOTA~\cite{Qu22} and SANGER~\cite{Lu21} can hardly extend to ReRAM-based sparse attention for the following reasons. 1) \textit{Vector-wise unmatched with matrix-wise.} DOTA and SANGER adopt vector-wise parallelism to utilize the spatial locality and reduce the random memory access, i.e., dividing the matrix into single vectors for processing. ReRAM naturally has matrix-wise parallelism (spatial locality failure). Thus, it is not feasible to map a vector-wise approach to matrix-wise hardware. 2) \textit{Less flexible matrix storage in ReRAM.} The input matrix $Q$ and $K$ can both be reordered and scheduled in DOTA. However, CPSAA can only reorder and schedule $Q$ because the $K$ matrix must be fixed in the crossbar arrays. The failure of scheduling $K$ makes the overall scheduling methods of DOTA and SANGER fail in ReRAM. 3) \textit{Difficult to remove all random memory access.} The SDDMM scheduling methods of DOTA and SANGER cannot eliminate the random memory access because they cannot access every non-zero value without access zero-values. Therefore, they can never achieve their ideal performance when extending to ReRAM. Three main challenges led to no available ReRAM-based SDDMM method, i.e., vectors binding, fixed $K$, and random memory access.}

Here we use the 4$\times$128 $Q$ and $K$ matrices as the example to explain our SDDMM acceleration method. As Figure~\ref{QKmeth} (b) shows, the elements of the $Q$ matrix are marked from `a' to `p', where `a' is a row vector with 32 32-bit numbers. {The most popular ReRAM-based mapping strategy (used in ReBERT and ReTransformer) of $K^\mathsf{T}$ uses 32 arrays to store each bit of 32-bit $K^\mathsf{T}$ (one array for one bit $K^\mathsf{T}$). This mapping binds all vectors of $K^\mathsf{T}$ and makes it impossible to utilize the vector-wise parallelism. So we map all bits of one vector into the same ReRAM array as Figure~\ref{QKmeth} (c) shows.} Each ReRAM array has 32 rows and 32 columns, and each ReRAM array stores 32 32-bit numbers, with each row storing one number. {We further adopt a ReCAM scheduler to generate the control signals for SDDMM operation, which can eliminate the random memory access of $Q$ as Figure~\ref{QKmeth} (a) shows. Although $K^\mathsf{T}$ is fixed in ReRAM arrays as shown in Figure~\ref{QKmeth} (e), our new mapping strategy can work well with the ReCAM scheduler to achieve a better scheduling results vs. DOTA. Next, we will introduce how our ReCAM scheduler controls the SDDMM operation.} 

{\bf ReRAM-ReCAM Coupling.} We assume that the 4$\times$4 matrix shown in Figure~\ref{QKmeth} (a) is the sparse mask matrix stored in a ReCAM array. In the beginning, the ReCAM array will perform the row-wise searching row-by-row as the colored arrows in Figure~\ref{QKmeth} (a) shows. We are taking the second row (green arrow line) in Figure~\ref{QKmeth} (a) as an example. The matching result of the second row indicates that the second row of $S$ has two zero elements (white cells), so we need to avoid the calculation of these zero-value cells. To locate the address of non-zero elements, the coordinates $\langle \alpha, \beta_i \rangle$ of these matched `1' cells will be sent to the CTRL. Then, the CTRL will utilize the row coordinate $\alpha$ to find the corresponding row of $Q$ as the green arrow in Figure~\ref{QKmeth} (b) shows. At the same time, the CTRL will utilize the column coordinate $\beta_i$ to find the corresponding ReRAM arrays in Figure~\ref{QKmeth} (e). As these green arrow lines in Figure~\ref{QKmeth} (e) show, the row indicated by $\alpha$ will be sent to the IR of these arrays indicated by $\beta_i$. The searching of the ReCAM array will iterate four times, and the other three iterations are marked with red arrow lines in Figure~\ref{QKmeth} (a). When all iterations are finished, the $Q$ matrix will be distributed to the IR of the ReRAM arrays, like these red and purple rectangles in Figure~\ref{QKmeth} (e) show. Finally, all arrays can perform the VMM operations using the topmost vector in their IR until all vectors are processed. To visually represent the advantage of our new approach, Figure~\ref{QKmeth} (d) uses the same color to present the elements which can be parallel calculated. We can see that our new method can finish the calculation of a 4$\times$4 $S$ matrix in two cycles.

{\bf Effect of Sparsity.} The sparsity of Figure~\ref{QKmeth} (a) is 0.5, so our novel method can reduce the latency from four cycles to two cycles. Note that the sparsity of the attention mechanism is around 0.1, so our approach can save up to 10$\times$ latency in real-world applications. In summary, our new method can finish the calculation of SDDMM with fewer latency iterations and higher parallelism. Further, our method can efficiently utilize the unstructured sparsity without complex {scheduler} and control signals.

\begin{figure}[t]
\centering
\includegraphics[width=8.9cm]{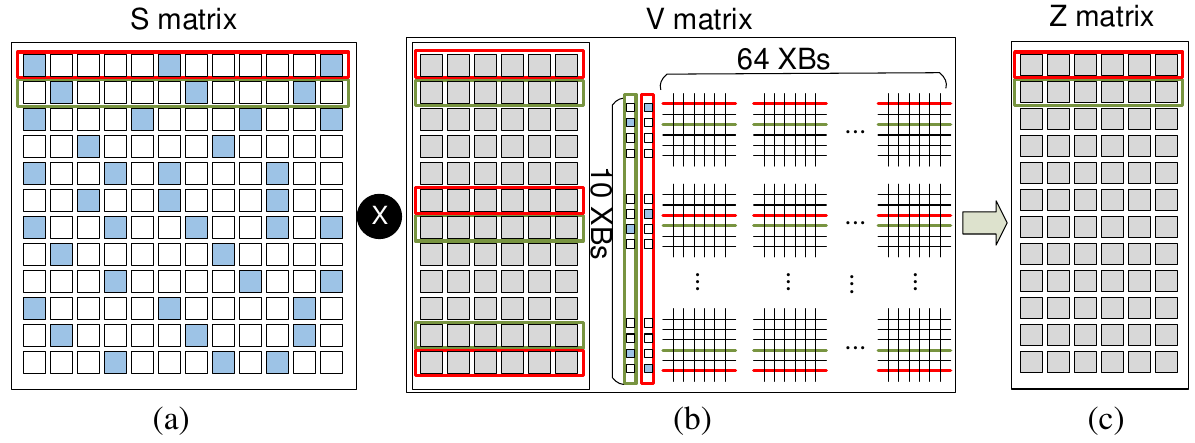}
\vspace{-2em}
\caption{The problem of current ReRAM-based SpMM method, (a) sparse $S$ matrix, (b) dense $V$ matrix, and (c) output $Z$ matrix}
\label{spmmpro}
\vspace{-1em}
\end{figure}
\subsection{SpMM Method}
\label{SpMM}
{\bf Current Problems.} {Many ReRAM-based SpMM solutions~\cite{Lin19, Song18} have been proposed. We classify them into two categories while extending to ReRAM-based sparse attention. One chooses to store sparse $S^\mathsf{T}$ in ReRAM while using dense $V^\mathsf{T}$ as the input. Another chooses to store the dense $V$ in ReRAM while using sparse $S$ as the input. A heavy wait-for-write latency occurs if we choose the first one. That is because we can not perform $Z = S\cdot V = (V^\mathsf{T}\cdot S^\mathsf{T})^\mathsf{T}$ until we finish the write of $S^\mathsf{T}$ (the write of $S^\mathsf{T}$ will become the critical datapath).} Figure~\ref{spmmpro} shows the visualization limitation of the second solution. As set in BERT~\cite{Devlin18} and A$^3$~\cite{Ham20}, $S$ is a 320$\times$320 sparse matrix and $V$ is a 320$\times$64 dense matrix. With the dense matrix $V$ stored in the ReRAM arrays (Figure~\ref{spmmpro} (b)), people can perform the VMM operation row-by-row with the sparse matrix $S$ as the input (Figure~\ref{spmmpro} (a)). Because many elements of $S$ are zero-value, current solutions choose to set zero signals for these zero values. However, these solutions will cause the array wasting problem, as Figure~\ref{spmmpro} (b) shows. Taking the first row of $S$ (marked in red frame rectangle) as an example, many rows (rows other than red) in ReRAM arrays keep idle. Therefore, the parallelism and {the runtime memory utilization} of ReRAM is quite low. In addition, the SpMM method in Figure~\ref{spmmpro} can only save energy but cannot save execution cycles.


\begin{figure}[t]
\centering
\includegraphics[width=8.9cm]{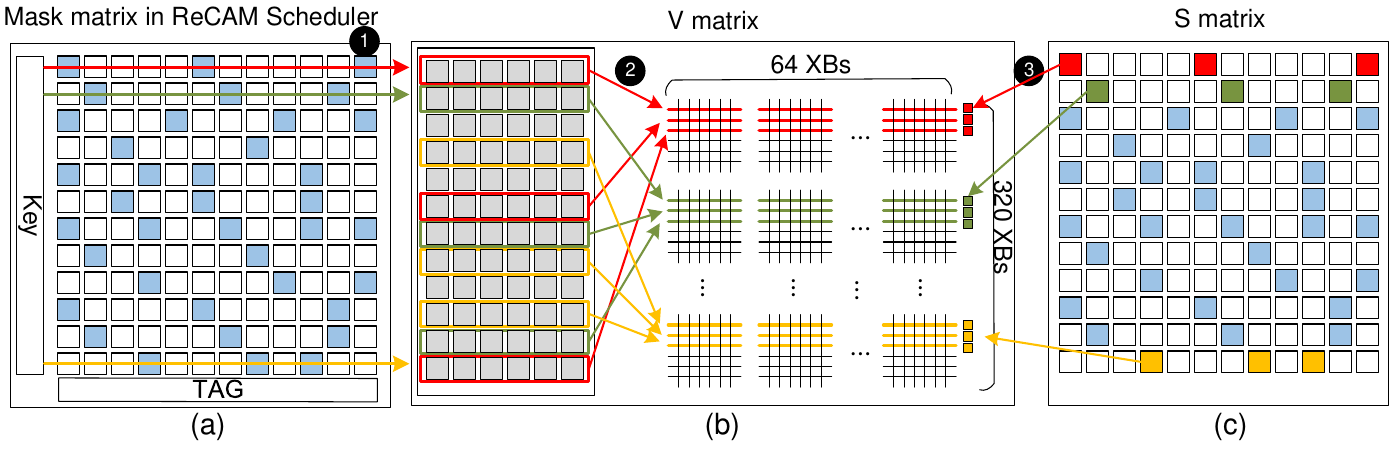}
\vspace{-2em}
\caption{Novel ReRAM-based SpMM design, (a) sparse mask matrix in the ReCAM Scheduler, (b) dense $V$ matrix, and (c) sparse $S$ matrix}
\label{SVmeth}
\vspace{-1em}
\end{figure}


{\bf Detailed Designs.} The sparsity of matrix $S$ is the same as the mask matrix, so it is possible to re-arrange the storage of the $V$ matrix using the mask matrix. Here we also use a 320$\times$320 sparse $S$ matrix and a 320$\times$64 dense $V$ matrix as an example to illustrate the details. As Figure~\ref{SVmeth} (a) shows, the mask matrix is stored in a ReCAM array (blue cells are `1' and white cells are `0'). First, the ReCAM array will perform a row-wise searching operation (red arrow line), which can find the coordinates $\langle \alpha, \beta_i \rangle$ of these matched `1' values (\circled{1}). Then, the coordinates of these matched cells will be sent to the CTRL, and the CTRL will use the column coordinates $\beta_i$ to find the corresponding rows of the $V$ matrix (red frame rectangle in Figure~\ref{SVmeth} (b)), which will be stored to the same ReRAM arrays (red lines in Figure~\ref{SVmeth} (b) \circled{2}). This mapping phase will iterate to each row of the ReCAM array, and we further show the mapping of the second row (green lines and rectangles) and the last row (yellow lines and rectangles). At the same time, the CTRL will also send the column coordinate $\beta_i$ to the sparse $S$ matrix. The $S$ matrix can find these elements (marked with red cells in Figure~\ref{SVmeth} (c)) corresponding to $\beta_i$ and send these cells to the corresponding IR (red cells in Figure~\ref{SVmeth} (b) \circled{3}). After all vectors are well mapped, CPSAA can simultaneously perform a VMM operation between the vectors in the IR and the $V$ matrix stored in the ReRAM array to calculate all rows of the output $Z$ matrix.

{\bf Advantages.} Assuming that the sparsity of the $S$ matrix is 0.1, so we need around 320$\times$64 32$\times$32 ReRAM arrays to store the matrix $V$. Meanwhile, the SpMM operation $Z = S\cdot V$ will be finished in one cycle. Following the traditional mapping shown in Figure~\ref{spmmpro}, 10$\times$64 32$\times$32 ReRAM arrays are required for storage $V$ matrix, and 320 cycles will be taken to generate the $Z$ matrix. The comparison shows that our novel SpMM method saves 320$\times$ execution time at the cost of 32$\times$ memory space. When memory space is limited, we can process the $S$ matrix in batches. In the case of two batches, we can get 160$\times$ execution time at the cost of 16$\times$ memory space.

\subsection{Processing Long Sequences}
\label{long}
This section describes the memory and time complexity of CPSAA when processing long sequences. We assume the input matrix $X$ has `T' tokens, and the dimension of each token are set to `D'.

{\bf Memory Complexity.} CPSAA has two types of memory, i.e., on-chip buffer and crossbar arrays. Four matrices are stored in the crossbar arrays, i.e., $W_S$, $W_V$, $X^\mathsf{T}$, and $V$. The memory complexity of $W_S$ and $W_V$ are $\mathbf{O}(D^2)$. The memory complexity of $X^\mathsf{T}$ and $V$ are $\mathbf{O}(D\times T)$. Four matrices are stored in the on-chip buffer, i.e., $X$, $M$, $S$, and $Z$. The memory complexity of $X$, $M$, and $Z$ are $\mathbf{O}(D\times T)$ while the memory complexity of $S$ is $\mathbf{O}(T^2)$. Therefore matrix $S$ is the only one that has quadratic memory complexity depending on sequence length `T'

{\bf Time Complexity.} In general, assume we are given two matrices A and B of sizes $\alpha \times \beta$ and $\beta \times \gamma$, respectively. The multiplication of $A*B$ requires $\alpha*\beta*\gamma$ multipliply-accumulate operations, and the result matrix $C = A*B$ is of size $\alpha \times \gamma$. Following the notations in Figure~\ref{compare} (c), we have $S = M \times X^\mathsf{T}$, where $M$ and $X$ are of sizes $T\times D$. Therefore, the time complexity of matrix $S$ is $\mathbf{O}(T^2)$. We also have $Z = S\times V$, and the sizes of $S$ and $V$ are $T\times T$ and $T\times D$, respectively. Therefore, the time complexity of matrix $Z$ is $\mathbf{O}(T\times D)$.

\subsection{Application-level Designs}
CPSAA is designed to eliminate the massive off-chip data transmissions in the current sparse attention accelerators. However, real-world NLP applications usually involve several attention layers working with {\em full-connection} (FC) layers, such as BERT~\cite{Devlin18}. The encoder is the basic computation unit of BERT, which contains one attention layer and several FC layer. We configure one CPSAA chip that works with several ReRAM-based FC layer to make up one encoder, and several encoders work together to make up the BERT-based solution.

We perform a pre-training and fine-tuning model to get the weight matrices, $W_Q$, $W_K$, and $W_V$, for a BERT-based text classification task~\cite{Devlin18}. Then, we pre-calculate $W_S = W_Q\cdot W_K^\mathsf{T}$ and ${\mathbf{QU}}(W_S)$. We pre-store the $W_S$, $W_V$, and ${\mathbf{QU}}(W_S)$ matrices in some ROAs. Note that all the above procedures are pre-processing because these matrices can be reused for different input sequences. In the beginning, an input matrix $X$ arrives at the {\em Input Buffer} (IB). First, the $X$ matrix will be processed by the CPSAA with no off-chip memory access to generate the $Z$ matrix. Second, the $Z$ matrix will be sent to the FC layer to generate the result of this encoder, which is processed following the ReRAM-based dot-product method proposed in ISAAC~\cite{Shafiee16}. Finally, the result of the previous encoder will be sent as the next encoder's input until the final inference results are obtained.

\begin{table}[t]
\centering
\tabcolsep=0.12cm
    \caption{CPSAA configurations}
    \small
    \renewcommand\arraystretch{1}
    \label{tab:breakdown}
    \tabcolsep=0.05cm
    \begin{tabular}{|C{1.66cm}||C{1.64cm}|c|c|c|}  
    \cline{1-5}
    
       {\bf Component} & {\bf Area (mm$^2$) } & {\bf Power (mW)} & {\bf Params.} & {\bf Spec.}\\
       \hline \hline
       
       \multicolumn{5}{|c|}{PCs properties} \\ \hline
       \multirow{3}*{\makecell{ReCAM \\ Scheduler}} & \multirow{3}*{0.0013} & \multirow{3}*{1.398} & {Bits per Cell} & {1} \\
       \cline{4-5}
       {} & {} & {} & {Size} & {512 $\times$ 512} \\
       \cline{4-5}
       {} & {} & {} & {Total} & {2} \\
       \hline
       {AIT} & {0.0608} & {36.89} & {Size} & {64KB} \\
       \hline
       {IB} & {0.0302} & {18.47} & {Size} & {32KB} \\
       \hline
       {CB} & {0.1217} & {74.21} & {Size} & {128KB} \\
       \hline
       {CTRL} & {0.0015} & {0.382} & {Total} & {1} \\
       \hline
       \multirow{2}*{SU} & \multirow{2}*{0.0072} & \multirow{2}*{1.134} & {LUT Size} & {512B} \\
       \cline{4-5}
       {} & {} & {} & {Total} & {1} \\
       \hline
       {QU\&DQU} & {0.0016} & {0.121} & {Total} & {1} \\
       \hline
       {{\bf PC} Total} & {0.2235} & {132.62} & {Size} & {288KB} \\
       \hline\hline
       
       \multicolumn{5}{|c|}{AG properties} \\ \hline
       \multirow{2}*{ADC} & \multirow{2}*{0.0015} & \multirow{2}*{2.0} & {Resolution} & {8 Bits} \\
       \cline{4-5}
       {} & {} & {} & {Total} & {1} \\
       \hline
       \multirow{3}*{XB Array} & \multirow{3}*{4.78e-5} & \multirow{3}*{0.581} & {Bits per Cell} & {1} \\
       \cline{4-5}
       {} & {} & {} & {Size} & {32 $\times$ 32} \\
       \cline{4-5}
       {} & {} & {} & {Total} & {12} \\
       \hline
       {S/H} & {4.69e-7} & {0.074} & {Total} & {12} \\
       \hline
       {DAC} & {6.38e-5} & {1.513} & {Total} & {32 $\times$ 12} \\
       \hline
       {IR} & {0.00049} & {0.294} & {Size} & {512B} \\
       \hline
       {OR} & {0.00036} & {0.108} & {Size} & {128B} \\
       \hline
       {S+A} & {0.00006} & {0.051} & {Total} & {1} \\
       \hline
       {{\bf AG} Total} & {0.00252} & {4.623} & {Size} & {2.1KB} \\
       \hline
        \multirow{2}*{\bf ROA} & \multirow{2}*{0.0278} & \multirow{2}*{50.87} & {AGs} & {11} \\
        \cline{4-5}
        {} & {} & {} & {Size} & {23.1KB} \\
        \hline
        \multirow{2}*{\bf WEA} & \multirow{2}*{0.1421} & \multirow{2}*{258.93} & {AGs} & {56} \\
        \cline{4-5}
        {} & {} & {} & {Size} & {117.6KB} \\
       
        
        \hline\hline
        \multicolumn{5}{|c|}{CPSAA properties} \\ \hline
        \multirow{2}*{\bf Tiles} & \multirow{2}*{25.18} & \multirow{2}*{28.32K} & {Total} & {64} \\
        \cline{4-5}
        {} & {} & {} & {Size} & {27.5MB} \\
        \hline
        {DTC} & {2.26} & {494.07} & {Total} & {1} \\
        \hline
        {{\bf CPSAA}} & {27.47} & {28.83K} & {Size} & {27.5MB} \\
        \hline
\end{tabular}
\vspace{-1em}
\end{table}

\section{Methodology}
\label{meth}
{\bf CPSAA Configurations.} We employ a Python cycle-accurate simulator to model the CPSAA attention accelerator. We use the 1000GB/s {\em On-Chip Interconnect} (OCI) for the on-chip transfer bandwidth~\cite{Jouppi2021}. ReRAM’s energy consumption is obtained with 7pJ per bit as in~\cite{Yan2020}. {As configured in ISAAC~\cite{Shafiee16}, the ``cycle" in CPSAA means the time of ADC processing 32 column signals, i.e., 25ns.}

The CPSAA configurations are shown in Table~\ref{tab:breakdown}. We configure CPSAA with 64 tiles, and each tile includes one group of peripheral components, 11 ROAs, and 56 WEAs. We use two 512$\times$512 ReCAM arrays as the Scheduler to store the mask matrices, while we use 32$\times$32 crossbar as the ReRAM array to perform the VMM operation. The crossbar configurations used in CPSAA are designed under the 32nm process with 533MHz clock frequency~\cite{Niu13}. Based TaOx ReRAM cells from~\cite{Niu13}, we conduct SPICE simulation for the area and power of crossbar configuration. We use CACTI 6.5~\cite{Muralimanohar09} in 32nm technology to evaluate the power and area of all registers. To obtain parameters of other peripheral components, we use Cadence-simulator~\cite{Krestinskaya15} for S/H, S+A in a crossbar. The power, area, and latency of a 2-bit precision DAC and a 2.0 mW 8b 1.0 GS/s ADC in 32nm CMOS are obtained from~\cite{Saberi11} and~\cite{Kull13}, respectively. The crossbars are read and written in a row parallel manner, and the SET/RESET latency is 1.52/2.11 ns for {\em single-level cells} (SLC)~\cite{Xu13}. The area and energy of SU, BU, QU, DQU, and CTRLs are established by SPICE circuits, too. The write endurance of the ReRAM cell can be alleviated greatly. For instance, considering up to 10$^{12}$ ReRAM write endurance~\cite{Zahoor20}, CPSAA can achieve hundreds of millions (10$^{8}$) of inferences when processing millions of tokens documents (10$^{4}$ times writing).


{\bf Benchmarks.} We use three typical attention-based NN models and get the weight matrices by pre-training, i.e., BERT-Base~\cite{Devlin18}, GPT-2~\cite{Radford19}, and BART~\cite{Lewis19}. BERT-Base (12 encoders) utilizes the encoder in Google Transformer to solve many NLP tasks. GPT-2 (12 decoders) uses the decoder of Transformer to process various NLP tasks. BART (six encoders and six decoders) is an optimized version of the Google Transformer. Our evaluation datasets include eight text classification tasks from the {\em General Language Understanding Evaluation} (GLUE) benchmark~\cite{Wang18glue} (CoLA, SST-2, MRPC, STS-B, QQP, MNLI, WNLI, RTE) and {\em Stanford Question Answering Dataset} (SQuAD) V2.0~\cite{Rajpurkar18}. 

\begin{figure}[t]
\centering
\includegraphics[width=8.9cm]{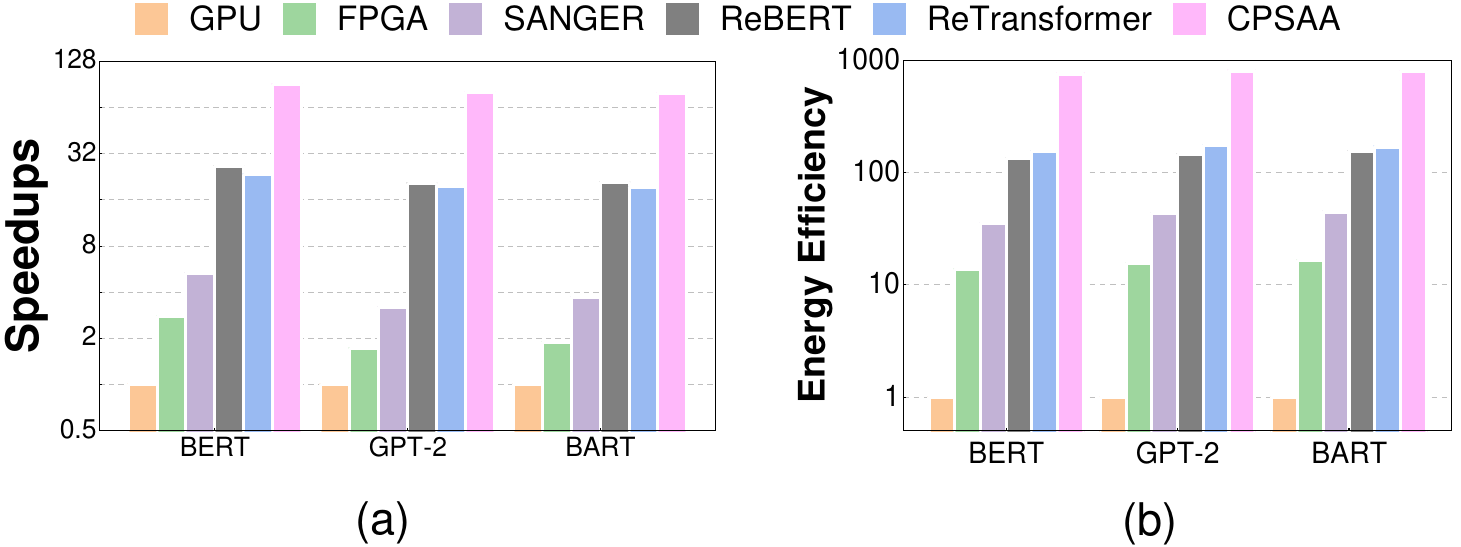}
\vspace{-2em}
\caption{(a) Average speedups of various models, (b) Average energy efficiency of various models}
\label{bart}
\vspace{-1em}
\end{figure}

{As configured in Transformer~\cite{Vaswani17}, BERT~\cite{Devlin18}, A$^3$~\cite{Ham20}, and SANGER~\cite{Lu21}, we set the model dimension d$_{model}$ = 512 and d$_K$ = d$_Q$ = 64 for all workloads. In addition, model dimension (d) is not likely to vary widely since a choice of too large d can lead to a decrease in model accuracy~\cite{Yin18}.} To fit the memory space of CPSAA, we divide all datasets into some little batches, and each batch has 320 embeddings, as set in BERT~\cite{Devlin18} and A$^3$~\cite{Ham20}. Embeddings in the same batch can be parallel processed by CPSAA without off-chip data transmission, and embeddings in different batches are processed in serial with small off-chip data transfers. We use the {\em Giga Operations per second} (GOPS) as the performance metric (throughput) and the {\em throughput per Watt} (GOPS/W) as the energy metric (energy efficiency).

{\bf Data Overflow Prevention.} The precision of input and weight matrices are set to 32-bit. The learning rate for GLUE and SQuAD v2.0 datasets are 2e-5 and 3e-5, and the number of fine-tuning epochs is set to 3 and 2, respectively. The attention mechanism involves lots of matrix multiplication calculations, so preventing data overflow is as important as ensuring accuracy. We extract the {\em Exponential Bit} (EB) of the whole array to make the {\em Fraction Bit} (FB) remaining 32-bit fixed-point as in~\cite{Feinberg18}. Thus, all VMM operations in CPSAA can be performed between 32-bit fixed-point numbers. After we get the results of the VMM operation between the FB, we can multiply the EB by the FB and get the final results.

{\bf Comparison Platforms.} CPSAA is compared with state-of-the-art platforms as follows.

We choose GPU as the NVIDIA TITAN RTX GPU@1770MHz, 576 Tensor cores, 4608 CUDA cores, 24GB memory, 672GB/s memory bandwidth, and 280 Watt TDP. We choose Bigbird as the attention algorithm for the GPU~\cite{Zaheer20}. We measure power consumption via nvidia-smi and execution latency using CUDA Event. {We measure the performance of GPUs using PyTorch with cuBLAS 11.2. 
We choose the FPGA-based software-hardware co-design accelerator proposed in~\cite{Zhang21} as the FPGA platform for comparison. The configurations are also following~\cite{Zhang21}.
We choose the configurations of SANGER as the ASIC-based attention accelerator for comparison~\cite{Lu21}.
We choose two architectures, ReBERT~\cite{Kang21} and ReTransformer~\cite{Yang20}, as the PIM-featured attention accelerator for comparison. The configurations of crossbars used in ReBERT and ReTransformer are the same as CPSAA.

\begin{figure}[t]
\centering
\includegraphics[width=8.9cm]{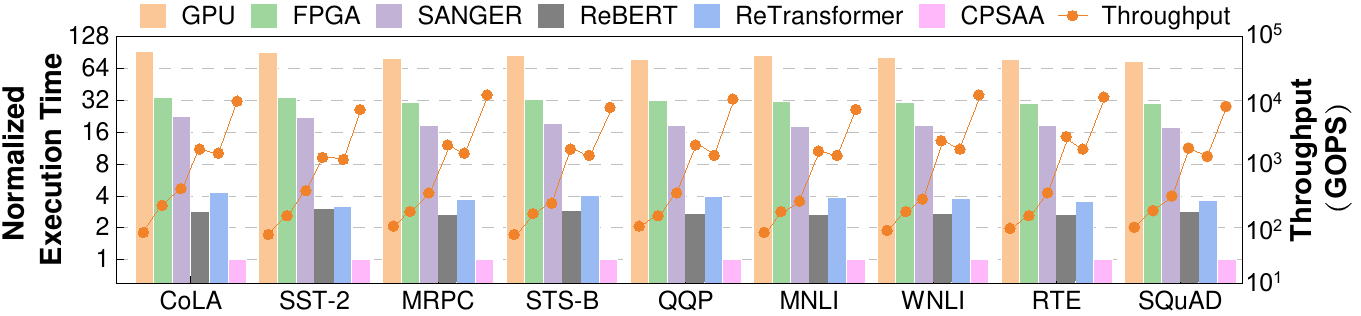}
\vspace{-2em}
\caption{Execution time normalized to CPSAA}
\label{kernel-p}
\vspace{-1em}
\end{figure}

\section{Evaluation}
\label{exper}
\subsection{Performance and Energy Efficiency}
Figure~\ref{bart} (a) and (b) present the average speedup and energy efficiency of different models, respectively. Because BERT, GPT-2, and BART are all based on encoders and decoders, while the current sparse attention accelerators can accelerate both the encoders and decoders. Therefore, the current sparse attention accelerators are effective in accelerating all models. The speedups of these three models have the same trend while the specific values vary by 10\%. To make the experimental results more concise, we present the results of BERT model to reveal the differences between CPSAA and other platforms. 

{\bf CPSAA against GPU and FPGA.} As shown in Figure~\ref{kernel-p}, the average throughput of GPU and FPGA platforms are 102 GOPS and 284 GOPS. Compared with the average 9142 GOPS of CPSAA, CPSAA has 89.6$\times$ and 32.2$\times$ performance improvement vs. the GPU and FPGA platforms, respectively. As for the energy efficiency shown in Figure~\ref{kernel-e}, GPU, FPGA, and CPSAA have an average of 0.63 GOPS/W, 8.6 GOPS/W, and 476 GOPS/W energy efficiency, respectively. Therefore, CPSAA can save 755.6$\times$ energy and 55.3$\times$ energy compared with GPU and FPGA platforms. CPSAA can achieve far better performance and energy-saving than GPU and FPGA platforms for the following reasons. First, CPSAA is a PIM-based platform, which can save lots of execution time by eliminating the massive off-chip random memory access. As shown in $\S~\ref{motivation}$, these off-chip data transmissions will take up to 60\% latency. Second, CPSAA elegantly exploits the high parallel VMM operation of ReRAM arrays by developing two sparse attention-specific methods. Compared with GPU- and FPGA-based platforms, these ReRAM-based matrix multiplication methods can achieve higher in-situ processing parallelism.


{\bf CPSAA against SANGER.} As Figure~\ref{kernel-p} and Figure~\ref{kernel-e} show, SANGER has an average of 513 GOPS throughput and 22.4 GOPS/W energy efficiency. Therefore, CPSAA has an average of 17.8$\times$ throughput improvement and 21.3$\times$ energy-saving compared with the ASIC-based platform SANGER. SANGER proposes a prediction-based pruning method to reduce the unnecessary calculations of the attention mechanism. SANGER also presents the ``splitting and packing" method to reduce the random memory access. The above optimizations make SANGER achieve 5.03$\times$ and 1.81 performance improvement compared with the GPU- and FPGA-based sparse attention accelerators. However, there are some unsolved issues left by SANGER, which CPSAA solves to achieve further speedups. First, SANGER adopts a software-based pruning method to generate its mask matrices, which has full data transfers between the memory and the off-chip processor. By proposing a PIM-based pruning method, CPSAA can generate the mask matrix while avoiding off-chip data transfers. Second, SANGER design the pruning phase to work serial with the attention calculation phase, while CPSAA can perform the pruning phase parallel with the attention calculation. Finally, CPSAA can directly and efficiently exploit unstructured sparsity by coupling ReCAM arrays to generate control signals for VMM operations, which can avoid the pretty high ``splitting and packing" overhead and the processing elements reconfiguration overhead in SANGER.

\begin{figure}[t]
\centering
\includegraphics[width=8.9cm]{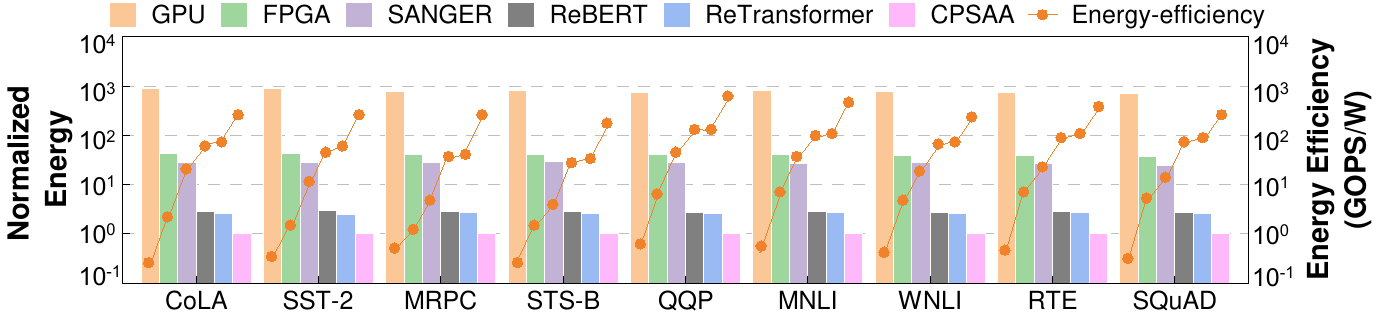}
\vspace{-2em}
\caption{Comsumed energy normalized to CPSAA}
\label{kernel-e}
\vspace{-1em}
\end{figure}


{\bf CPSAA against ReBERT and ReTransformer.} The throughput and energy efficiency of ReBERT and ReTransformer are also shown in Figure~\ref{kernel-p} and Figure~\ref{kernel-e}. ReBERT and ReTransformer have 2696 GOPS and 2381 GOPS throughput, respectively. Therefore, CPSAA achieves 3.39$\times$ and 3.84$\times$ performance improvement vs. ReBERT and ReTransformer. ReBERT and ReTransformer have an average of 83.7 GOPS/W and 97.1 GOPS/W energy efficiency, respectively. Thus, CPSAA has an average of 5.7$\times$ and 4.9$\times$ energy saving compared with ReBERT and ReTransformer. We also analyze the reasons for these results as follows. ReBERT and ReTransformer are both PIM-based platforms, and they can achieve performance improvement against the GPU- and FPGA-based platforms since they do not contain the off-chip data transfers. Compared with ReBERT and ReTransformer, CPSAA designs efficient ReRAM-based SDDMM and SpMM methods, which can save lots of execution time by avoiding these unnecessary VMM operations. Moreover, the novel attention calculation mode of CPSAA can hide the pretty high ReRAM write overhead while maintaining the VMM parallelism. Although CPSAA introduces an extra sparsity pruning phase, it can work parallel with the attention calculation and not introduce extra latency.

\begin{figure}[t]
\centering
\includegraphics[width=8.3cm]{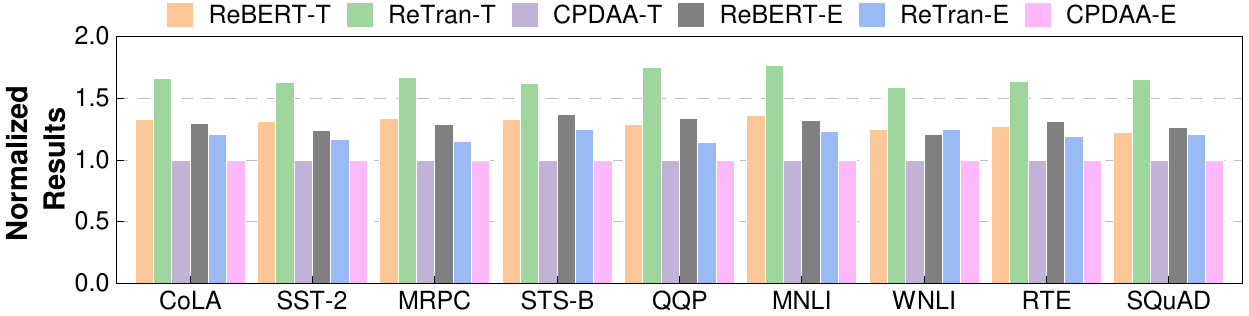}
\vspace{-1em}
\caption{Compared CPDAA with ReBERT and ReTransformer, which is normalized to CPDAA}
\label{cpdaa}
\vspace{-1em}
\end{figure}

\begin{figure}[t]
\centering
\includegraphics[width=8.3cm]{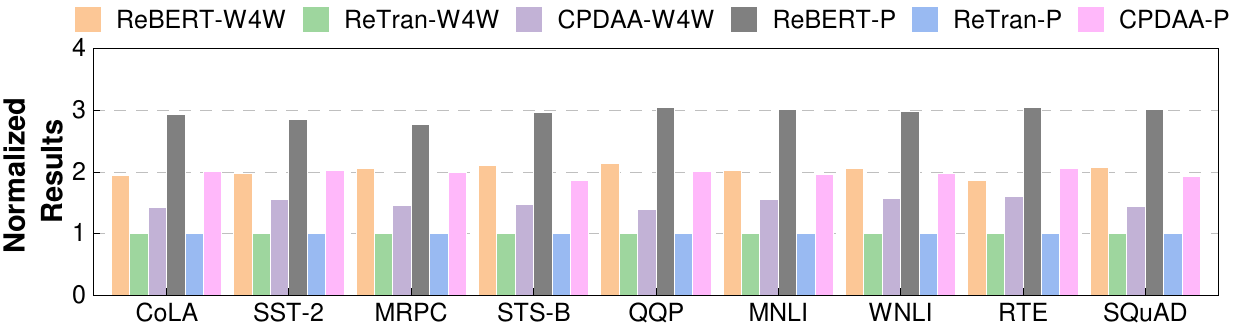}
\vspace{-1em}
\caption{Compared CPDAA with ReBERT and ReTransformer, which is normalized to ReTransformer}
\label{Asy}
\vspace{-1em}
\end{figure}

\subsection{Effectiveness of Calculation Mode}
This work proposes a new calculation mode to hide the ReRAM write overhead {while improving parallelism by removing data dependency}. Therefore, we also design experiments to evaluate the effectiveness of our calculation mode, as shown in Figure~\ref{cpdaa}. We design a {\em dense-version of CPSAA}, called CPDAA, to eliminate the acceleration effect of the sparsity. Figure~\ref{cpdaa} shows the execution time and energy consumption of ReBERT, ReTransformer, and CPDAA, which is normalized to {\em the execution time of CPDAA} (CPDAA-T) and {\em the energy consumption of CPDAA} (CPDAA-E). It is shown that ReBERT and ReTransformer take 1.31$\times$ and 1.64$\times$ execution time against CPDAA, respectively. As for the energy consumption, ReBERT and ReTransformer consume 1.30$\times$ and 1.21$\times$ than CPDAA. To further reveal the reasons for these results, we design more experiments as follows.

We further design two metrics to compare the calculation mode of CPDAA with ReBERT and ReTransformer, i.e., the {\em execution time of waiting for write} (CPDAA-W4W) and the {\em number of arrays for parallel VMM operation} (CPDAA-P). The experimental results in Figure~\ref{Asy} are all normalized to ReTransformer (ReTran-W4W and ReTran-P). ReBERT and CPDAA take 1.94$\times$ and 1.48$\times$ execution time on write operations compared with ReTransformer. In contrast, the VMM parallelism of ReBERT and CPDAA is 2.88$\times$ and 2.03$\times$ compared with ReTransformer. ReTransformer has the minimal write latency but the worst VMM parallelism because it is designed to reduce the write overhead as much as possible. However, they turn the VMM operations that could be executed in parallel into strictly serial execution, reducing the wait time for write operations while increasing the wait time for the previous VMM operations. ReBERT has the maximal write overhead but the best VMM parallelism because they use a write-then-calculate computational mode, which maximizes the VMM execution efficiency but takes longer to wait for write. CPDAA takes a trade-off between ReBERT and ReTransformer, which reduces unnecessary write overhead while preserving necessary VMM parallelism.

\begin{figure}[t]
\centering
\includegraphics[width=8.3cm]{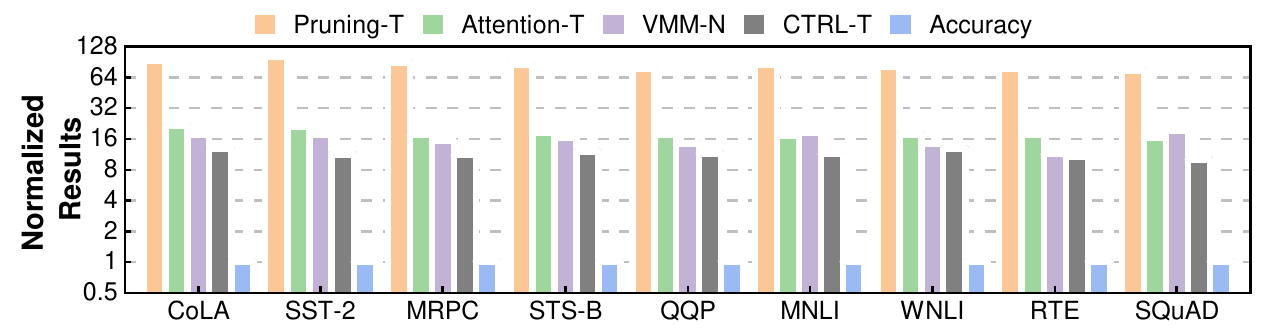}
\vspace{-1em}
\caption{Compared the pruning method of CPSAA with SANGER}
\label{sanger_com}
\vspace{-1em}
\end{figure}

In the experiments shown in Figure~\ref{Asy}, ReTransformer optimizes the write operations, but the performance of which is worse than ReBERT for the following reason. Taking into account the non-negligible write overhead, this work designs all ReRAM-based platforms to use a {\em single-level cell} (SLC) with lower write latency and lower write energy consumption. If a {\em multi-level cell} (MLC) with higher write overhead is used, the advantage of ReTransformer will be revealed.

\subsection{Acceleration of Pruning Architecture}
CPSAA proposes a novel ReRAM-based PIM-featured pruning method to generate the mask matrix. Therefore, we also develop experiments to evaluate the effectiveness of our novel pruning architecture. As Figure~\ref{sanger_com} shows, we choose five indicators to reveal the advantages of our novel method against the state-of-the-art pruning method proposed in SANGER. They are the {\em execution time of the pruning phase} (Pruning-T), the {\em execution time of the attention calculation phase} (Attention-T), the {\em number of pruning phase's VMM operations} (VMM-N), the {\em {CTRL runtime scheduling time}} (CTRL-T), and the {\em accuracy comparison} (Accuracy). The results in Figure~\ref{sanger_com} are all normalized to CPSAA, i.e., the multiple of SANGER divided by CPSAA.

The pruning phase in SANGER takes 85.1$\times$ execution time compared with our novel pruning method. That is because our novel method eliminates the off-chip data transfers and greatly improves the computational parallelism when generating mask matrices, which can save lots of execution times. Moreover, the Attention-T of SANGER is 18.7$\times$ of CPSAA, which comes from two aspects. First, CPSAA eliminates massive off-chip data transfers compared with SANGER. Second, CPSAA can leverage the high parallel VMM operations of the ReRAM arrays, which can greatly reduce the latency of the VMM operations. Figure~\ref{sanger_com} also shows that CPSAA can save around 16.37$\times$ VMM operations when generating the mask matrix. CPSAA can generate the mask matrix directly using the input matrices, while SANGER needs lots of VMM operations to generate some intermediate matrices $Q$ and $K$ first. 

Experimental results show that SANGER takes 11.4$\times$ execution time on CTRL-T when compared with CPSAA. That is because SANGER's ``splitting and packing" algorithm has complex scheduling control signals for sparse $S$ and pretty high processing elements re-configuration overhead. CPSAA can use ReCAM Scheduler to reduce the control signals and eliminate the re-configuration overhead by designing a ReRAM-ReCAM coupling SDDMM and SpMM methods. CPSAA also adopts a new equation using the low-precision weight matrix $W_S$ and quantified $M$ to generate the mask matrix. Therefore, it is necessary to evaluate the accuracy of our pruning method. Figure~\ref{sanger_com} shows that CPSAA loses less than 0.2\% of accuracy compared to SANGER. 

\begin{figure}[t]
\centering
\includegraphics[width=8.3cm]{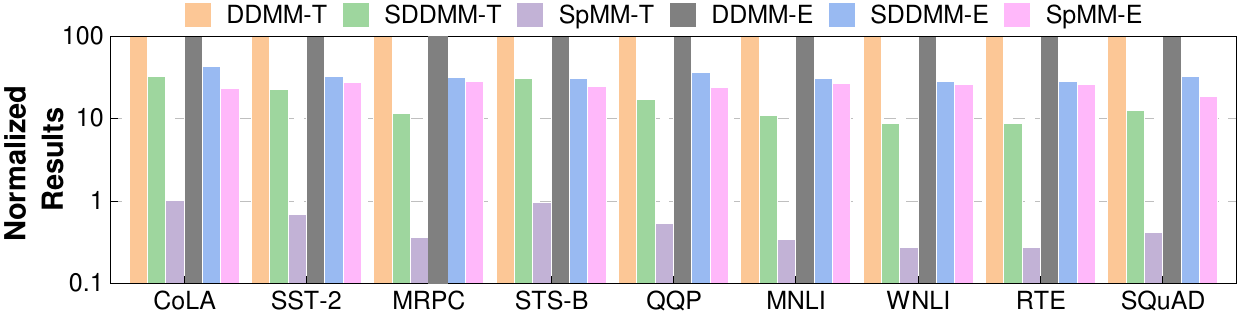}
\vspace{-1em}
\caption{Compared the SDDMM and SpMM methods with the DDMM methods in ReBERT, which is normalized to the DDMM operation}
\label{Sparse}
\vspace{-1em}
\end{figure}

\subsection{Novel SDDMM and SpMM Designs}
Compared to current PIM-based dense attention accelerators ReBERT and ReTransformer, CPSAA designs two novel ReRAM-based methods to accelerate the SDDMM and SpMM operations, respectively. Therefore, we also design experiments to reveal the effectiveness of the novel SDDMM and SpMM methods. As shown in Figure~\ref{Sparse}, the metrics of this evaluation are the execution time and the consumed energy. The baseline is the {\em execution time of the DDMM operations} (DDMM-T) and the {\em consumed energy of the DDMM operations} (DDMM-E) in ReBERT. All experimental results are normalized to DDMM-T (100) and DDMM-E (100). The latency of the SDDMM and the SpMM methods are 17.5\% and 0.54\% of the DDMM method, respectively. The reason for the speedups of the SDDMM operation is as follows. The SDDMM approach can minimize these unnecessary VMM calculations and greatly reduce the latency by using ReCAM Scheduler to guide the generation of control signals. Our new SpMM method can significantly reduce the number of idle rows in VMM operations, greatly increasing the parallelism and reducing the latency. Figure~\ref{Sparse} also shows that the consumed energy of the novel SDDMM and SpMM methods are 32.9\% and 25.2\% of the DDMM method. It is easy to understand that the SDDMM method can save energy by using the mask matrix to avoid computing unnecessary VMM operations. The novel SpMM method can save energy by avoiding a large number of idle rows of the ReRAM array.

\begin{figure}[t]
\centering
\subfloat[]{
\begin{minipage}[t]{0.489\linewidth}
\centering
\includegraphics[width=1.55in]{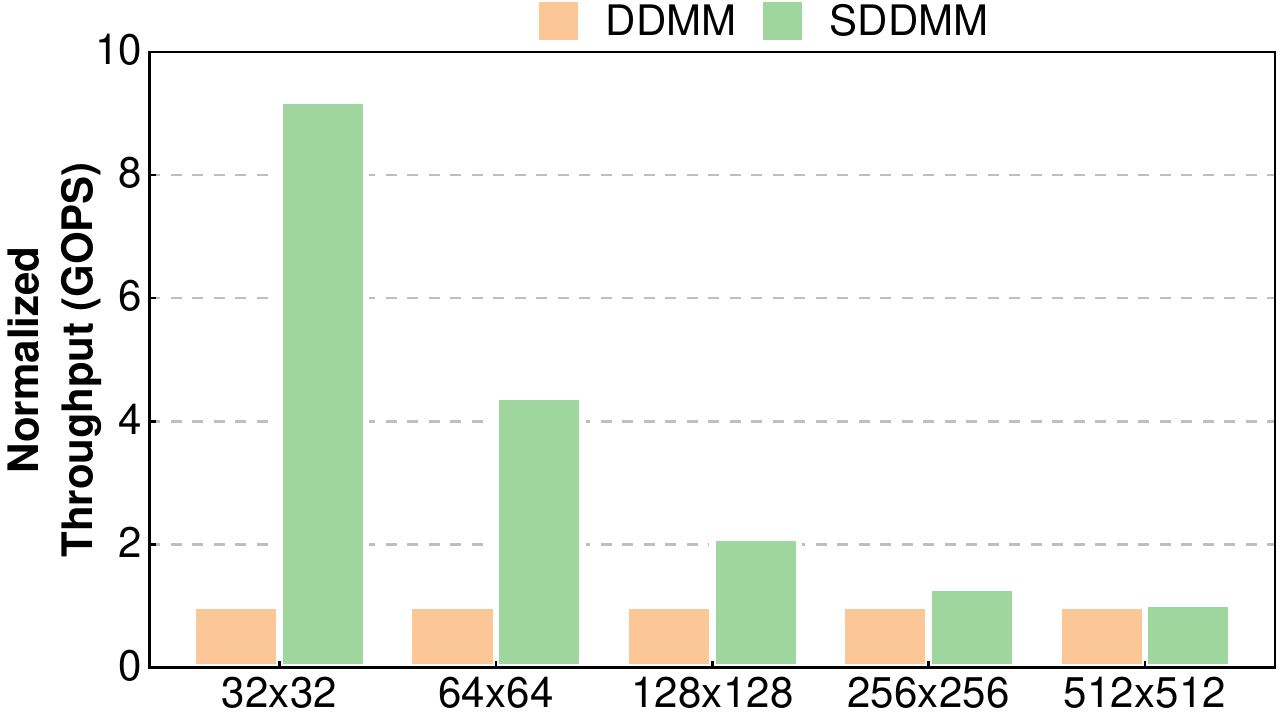}
\end{minipage}
}%
\subfloat[]{
\begin{minipage}[t]{0.489\linewidth}
\centering
\includegraphics[width=1.58in]{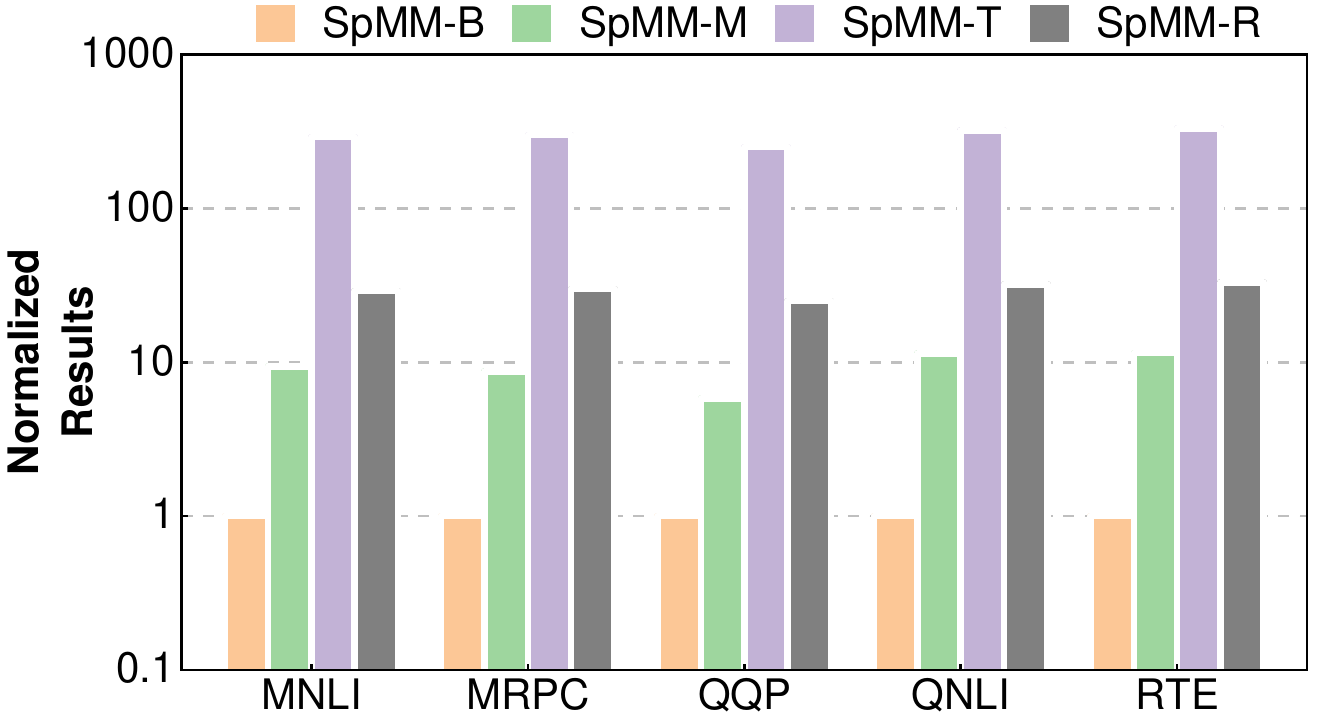}
\end{minipage}
}%
\centering
\caption{(a) Throughput with different size of crossbar, (b) compared the SpMM method in Figure~\ref{spmmpro} with our SpMM method in Figure~\ref{SVmeth}}
\vspace{-1em}
\label{size}
\end{figure}

{We also design experiments to reveal how the proposed SDDMM performance is affected by the crossbar size, as Figure~\ref{size} (a) shows. The X-axis indicates the crossbar size, and the Y-axis indicates the average (average of running all datasets) speedup of our SDDMM method vs. the ReRAM-based DDMM solution. We can find that the speedups of the SDDMM method decrease as crossbar size increases. That is because the key idea of our SDDMM method is to convert the matrix-wise parallelism of the ReRAM array to vector-wise parallelism. As the crossbar size arise, more vectors can be stored in the same array, and the vector-wise parallelism will decrease. Based on the above phenomena, we recommend using an array size that matches the value precision to maximize vector-wise parallelism.}

{In Figure~\ref{size} (b), we design experiments to reveal the runtime memory utilization, throughput, and data replication between the SpMM method in Figure~\ref{spmmpro} and Figure~\ref{SVmeth}. SpMM-B refers to the baseline SpMM method in Figure~\ref{spmmpro} and all its runtime memory utilization, throughput, and data replication are normalized to 1. SpMM-M, SpMM-T, and SpMM-R refer to the runtime memory utilization, throughput, and data replication of our SpMM method. For all five datasets, our SpMM method has an average of 9.36$\times$ runtime memory utilization improvement, 298$\times$ throughput improvement, and 30.4$\times$ data replication overhead. These experimental results are consistent with the CPSAA design principle of using space cost in exchange for low SpMM latency.}


\begin{figure}[t]
\centering
\subfloat[]{
\begin{minipage}[t]{0.489\linewidth}
\centering
\includegraphics[width=1.65in]{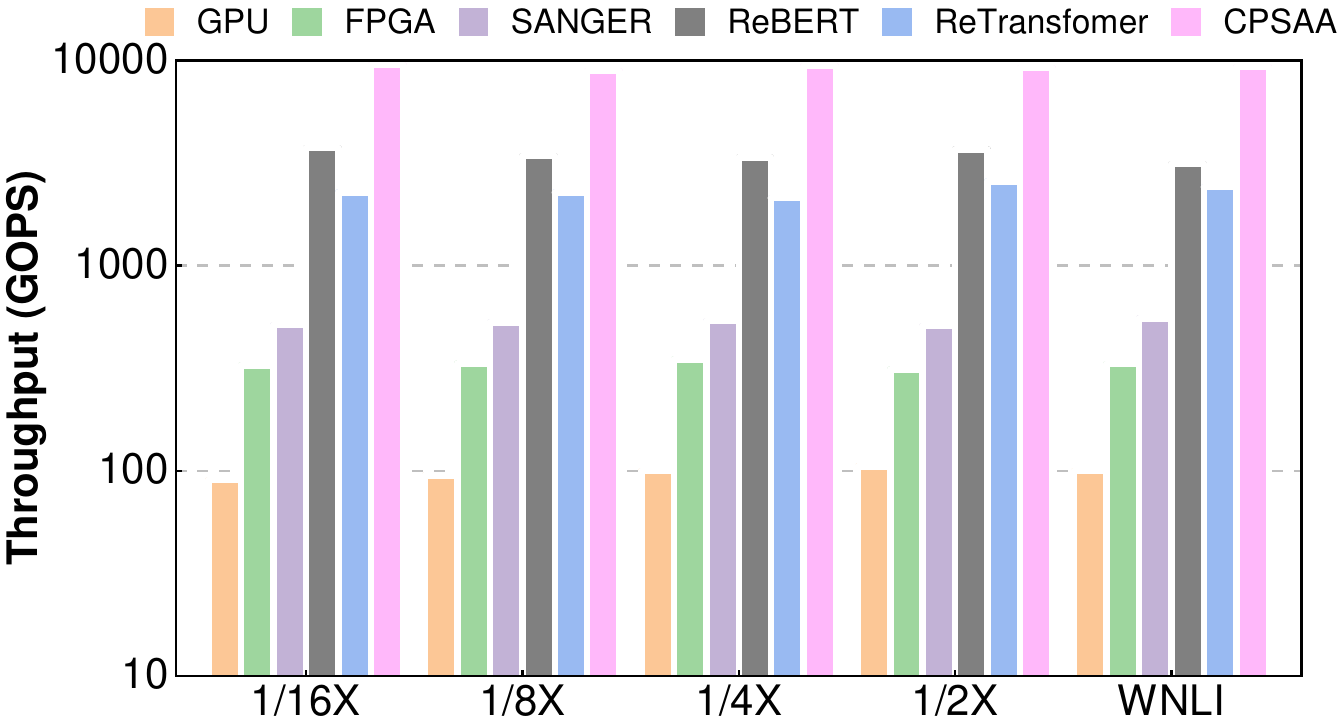}
\end{minipage}
}%
\subfloat[]{
\begin{minipage}[t]{0.489\linewidth}
\centering
\includegraphics[width=1.55in]{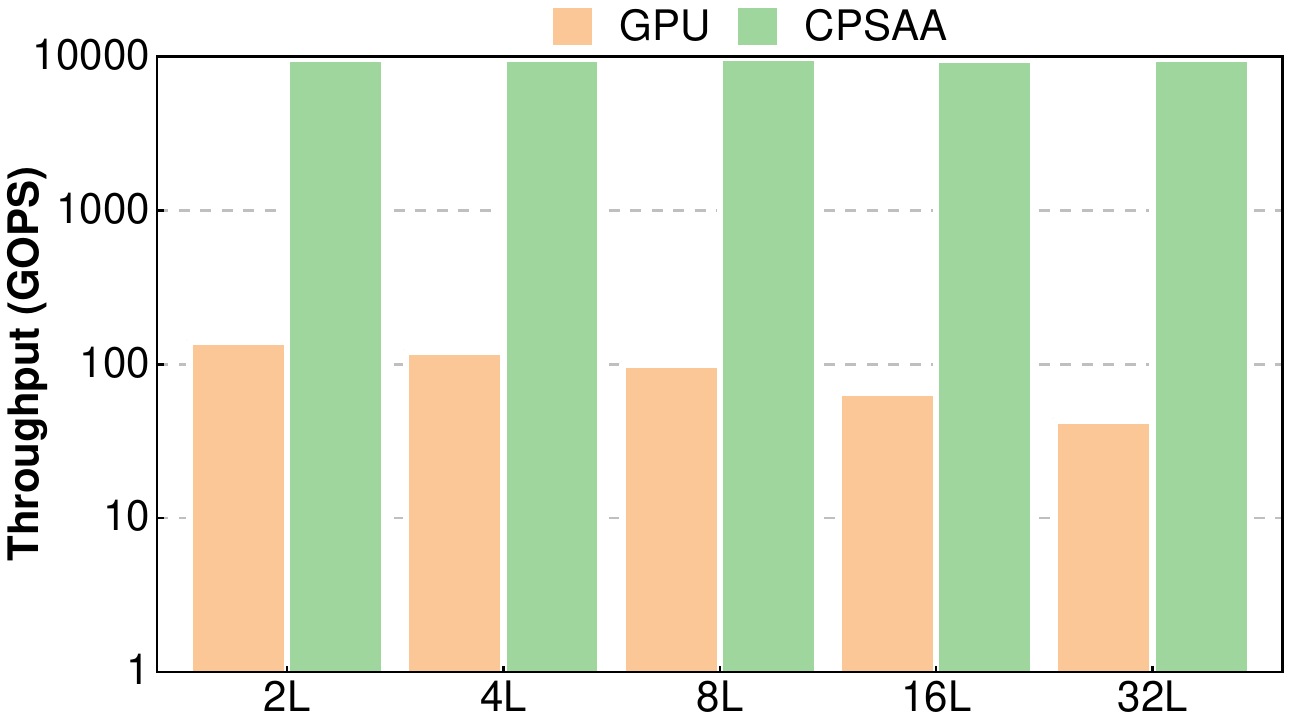}
\end{minipage}
}%
\centering
\caption{(a) Throughput with different size of datasets, (b) throughput with different encoder layers}
\label{scalability}
\vspace{-1em}
\end{figure}

\begin{figure}[t]
\centering
\includegraphics[width=8.3cm]{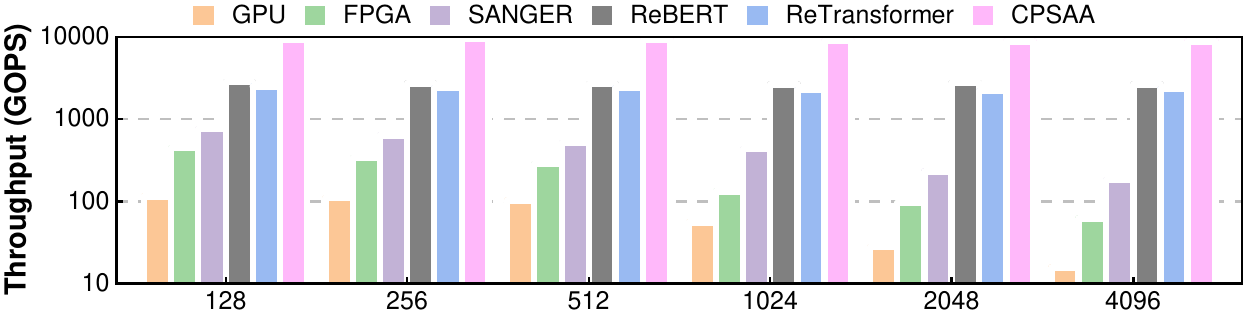}
\vspace{-1em}
\caption{Throughput with different sequence lengths}
\label{length}
\vspace{-1em}
\end{figure}

\subsection{Scalability}
First, we study the scalability concerning the data volume of input embeddings on the WNLI dataset. Because we cannot get datasets with multiplied data volumes in real-world datasets, we randomly select 1/16 to 1/2 input embeddings in the WNLI dataset to evaluate the effect of the data volume. Figure~\ref{scalability} (a) shows the throughput of GPU, FPGA, SANGER, ReBERT, ReTransformer, and CPSAA processing various sizes of the WNLI dataset. The throughput of CPSAA can be maintained at a relatively stable value. That is because we divide the input embeddings into small batches with 320 sequences, and different batches are processed serially. Therefore, a larger dataset has more batches and takes more execution time, making the number of {\em Giga operations per second} (GOPS) remain stable.

We study the scalability concerning the number of encoder layers in BERT. We set the number of the encoder from two layers (2L) to 32 layers (32L), and we choose the GPU baseline as the comparison platform. As Figure~\ref{scalability} (b) shows, the throughput of GPU-based platforms declines noticeably when the number of encoder layers increases. Unlike the trend in the GPU platform, the throughput of CPSAA remains stable. That is because more encoder layers will introduce more computational tasks, generate more intermediate matrices, and require more memory space. Therefore, the GPU-based platform has more random memory access while taking more execution time. But in the PIM-based platform such as CPSAA, more memory space means more computational resources, and the throughput of CPSAA will not increase as the number of encoder layers increases.

We design experiments to reveal the sequence lengths scalability. We configure six sequence lengths, i.e., 128, 256, 512, 1024, 2048, and 4096. Figure \ref{length} shows the average throughput of running all datasets. The three von Neumann accelerators, GPU, FPGA, and SANGER, show the same trend of sequence length scalability. The CPU-, FPGA- and SANGER-based accelerators can maintain stable throughput when the sequence lengths $\leq$512. When the sequence lengths are greater than 512, the throughput of the above platforms decreases significantly as the sequence lengths increase. That is because the increased sequence length generates a large amount of intermediate data, which exacerbates the random access and off-chip transfers of the von Neumann accelerators, thereby reducing system efficiency. The above results are consistent with the time complexity analysis in Section~\ref{long}. The throughput of the PIM-based accelerators, ReBERT, ReTransformer, and CPSAA, do not vary significantly with sequence lengths. That is because the PIM-based accelerators can on-chip process all the intermediate data without transferring them to the main processor.

\section{conclusion}
\label{concl}
We investigate current sparse attention accelerators and find the reasons for their limited speedups. This work presents a crossbar-based PIM-featured sparse attention accelerator, CPSAA, to tackle current problems. We design a novel calculation mode of attention mechanism to hide the write overhead while maintaining the VMM parallelism. We also present a new PIM-based pruning method to eliminate the off-chip memory access when generating the mask matrix. To solve the challenges that all current ReRAM-based attention accelerators can hardly efficiently extend to sparse attention, we design novel ReRAM-based SDDMM and SpMM methods. Finally, we design experiments to evaluate the effectiveness of all designs proposed in this work. Our experimental results show that in performance and energy-saving, CPSAA outperforms state-of-the-art accelerators, ranging from GPU, FPGA, ASIC, and PIM. The experimental results also reveal the effectiveness of all these core designs mentioned in this work.


\bibliographystyle{IEEEtranS}
\bibliography{refs}

\begin{IEEEbiography}[{\includegraphics[width=1in,height=1.25in,clip,keepaspectratio]{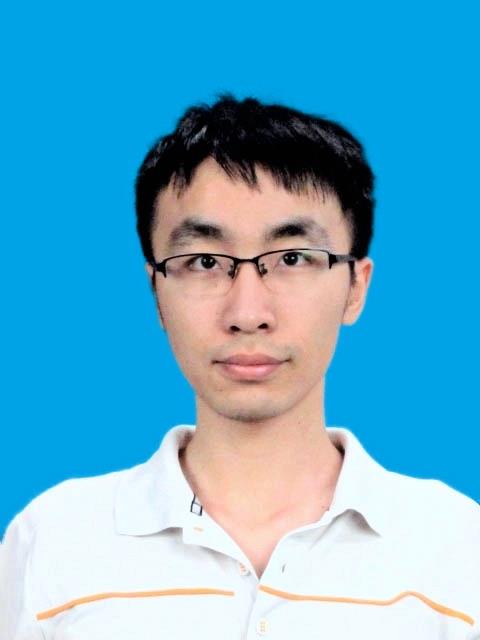}}]{Hui-Ze Li}
received his Ph.D. degree from the School of Computer Science and Technology, Huazhong University of Science and Technology (HUST), in 2022. He is now working as a Research Fellow in School of Computing, National University of Singapore (NUS). His current research interests include computer architecture, emerging non-volatile memory, and processing in memory.
\end{IEEEbiography}

\begin{IEEEbiography}[{\includegraphics[width=1in,height=1.25in,clip,keepaspectratio]{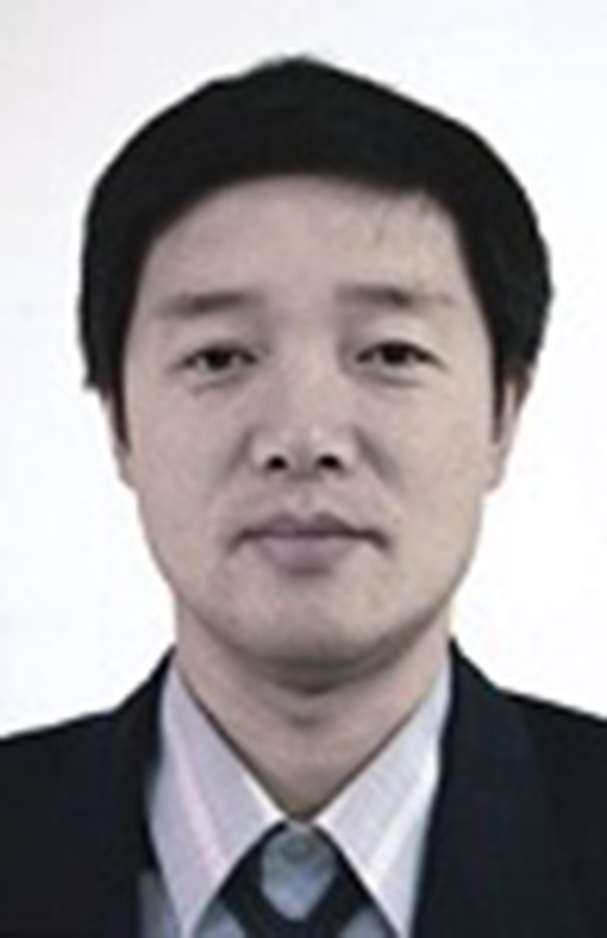}}]{Hai Jin}
is a Chair Professor of computer science and engineering at Huazhong University of Science and Technology (HUST) in China. Jin received his PhD in computer engineering from HUST in 1994. In 1996, he was awarded a German Academic Exchange Service fellowship to visit the Technical University of Chemnitz in Germany. Jin worked at The University of Hong Kong between 1998 and 2000, and as a visiting scholar at the University of Southern California between 1999 and 2000. He was awarded Excellent Youth Award from the National Science Foundation of China in 2001. 
Jin is a Fellow of IEEE, Fellow of CCF, and a life member of the ACM. He has co-authored more than 20 books and published over 900 research papers. His research interests include computer architecture, parallel and distributed computing, big data processing, data storage, and system security.
\end{IEEEbiography}

\begin{IEEEbiography}[{\includegraphics[width=1in,height=1.25in,clip,keepaspectratio]{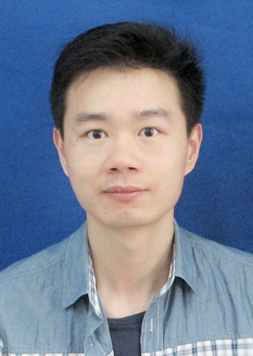}}]{Long Zheng}
is now an associate professor in the School of Computer Science and Technology, Huazhong University of Science and Technology (HUST), Wuhan. He received his Ph.D. degree in computer engineering at HUST, Wuhan, in 2016. His current research interests include program analysis, runtime systems, and configurable computer architecture with a particular focus on graph processing.
\end{IEEEbiography}

\begin{IEEEbiography}[{\includegraphics[width=1in,height=1.25in,clip,keepaspectratio]{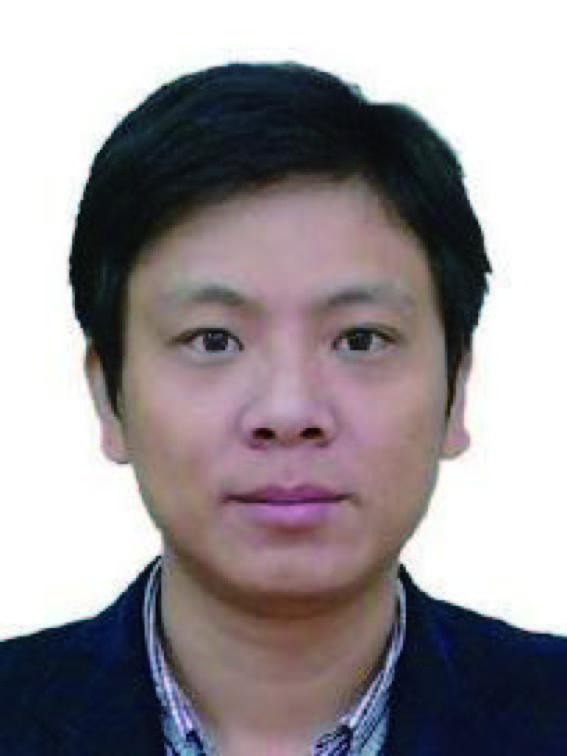}}]{Xiao-Fei Liao }
received his Ph.D. degree in computer science and engineering from Huazhong University of Science and Technology (HUST), Wuhan, in 2005. He is now a Professor in the School of Computer Science and Technology at HUST, Wuhan. He has served as a reviewer for many conferences and journal papers. His research interests are in the areas of system software, P2P system, cluster computing and streaming services. He is a member of IEEE and the IEEE Computer Society.
\end{IEEEbiography}

\begin{IEEEbiography}[{\includegraphics[width=1in,height=1.25in,clip,keepaspectratio]{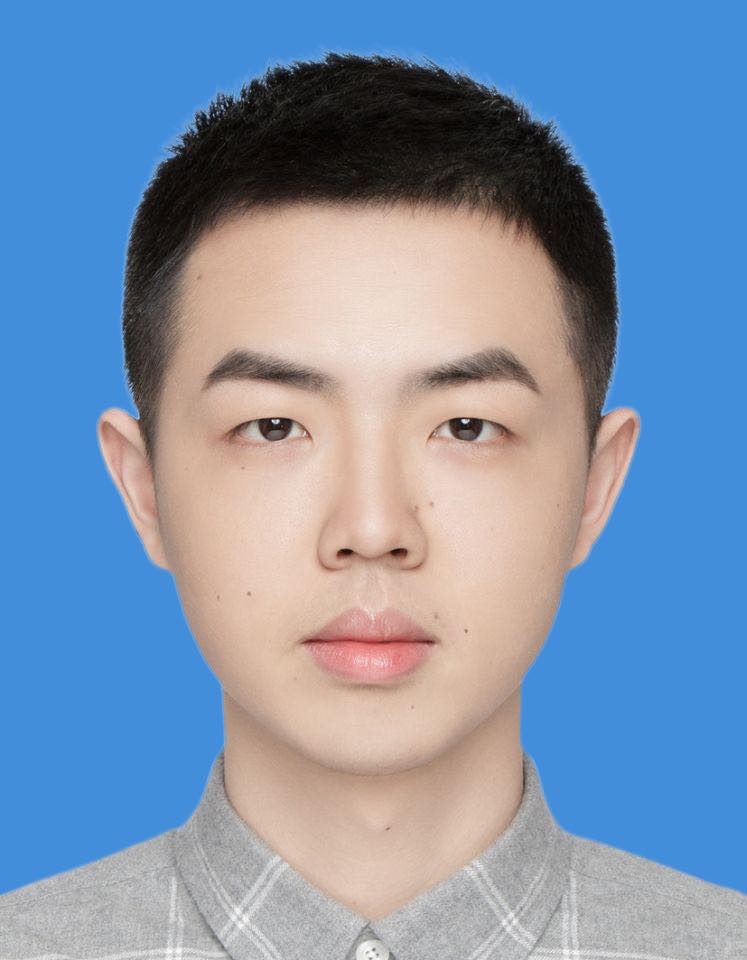}}]{Yu Huang}
received his Ph.D. degree from the Huazhong University of Science and Technology (HUST), in 2022. He is now working as a postdoc in the School of Computer Science and Technology, HUST, in China. His research interests focus on processing-in-memory architecture and graph processing.
\end{IEEEbiography}

\begin{IEEEbiography}[{\includegraphics[width=1in,height=1.25in,clip,keepaspectratio]{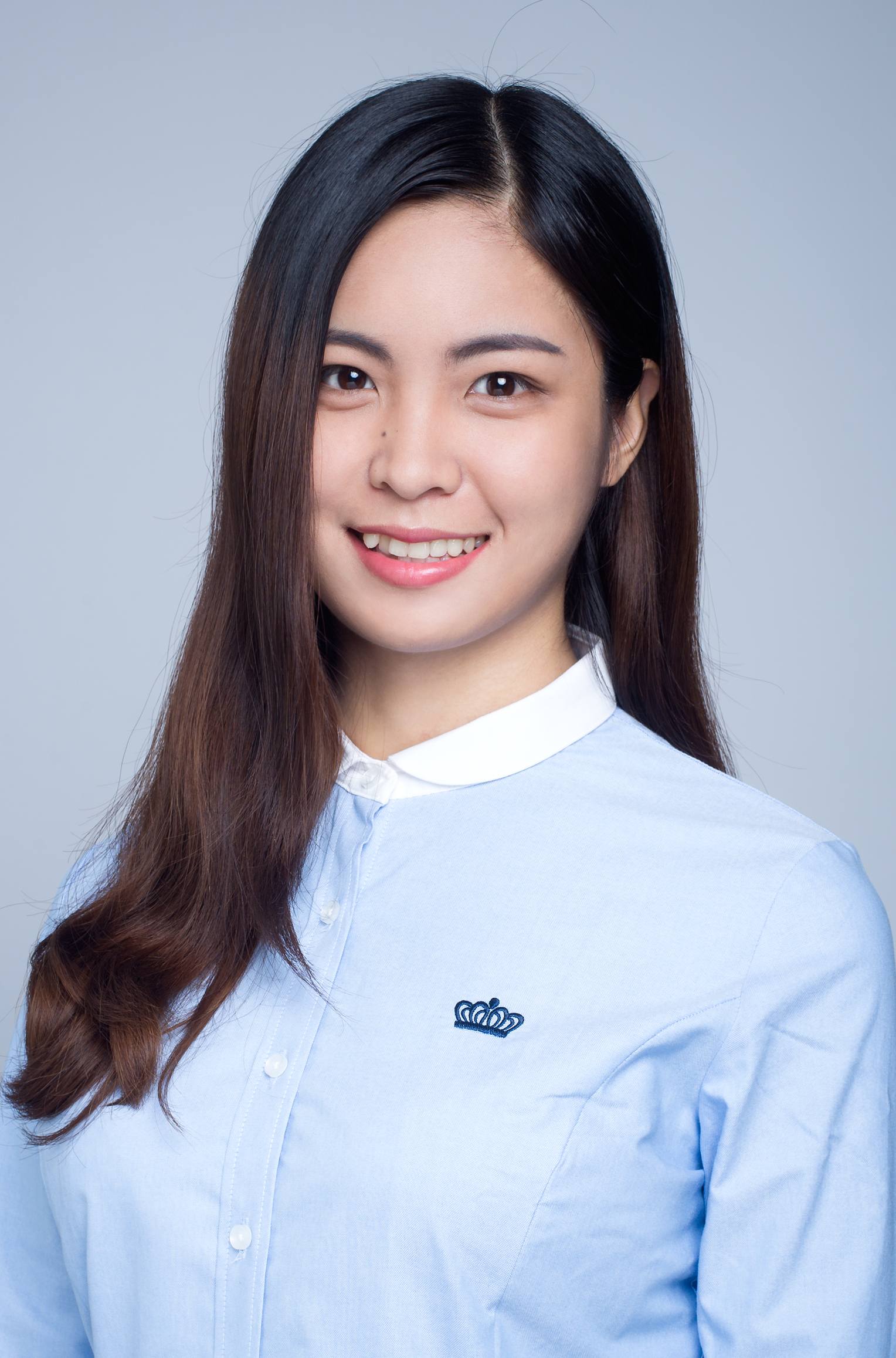}}]{Cong Liu}
received the bachelor's degree from Dalian Maritime University, China, in 2018. She is currently working toward the PhD degree with the Huazhong University of Science and Technology in China. Her research interests include in-memory computing and ReRAM-based accelerator.
\end{IEEEbiography}

\begin{IEEEbiography}[{\includegraphics[width=1in,height=1.25in,clip,keepaspectratio]{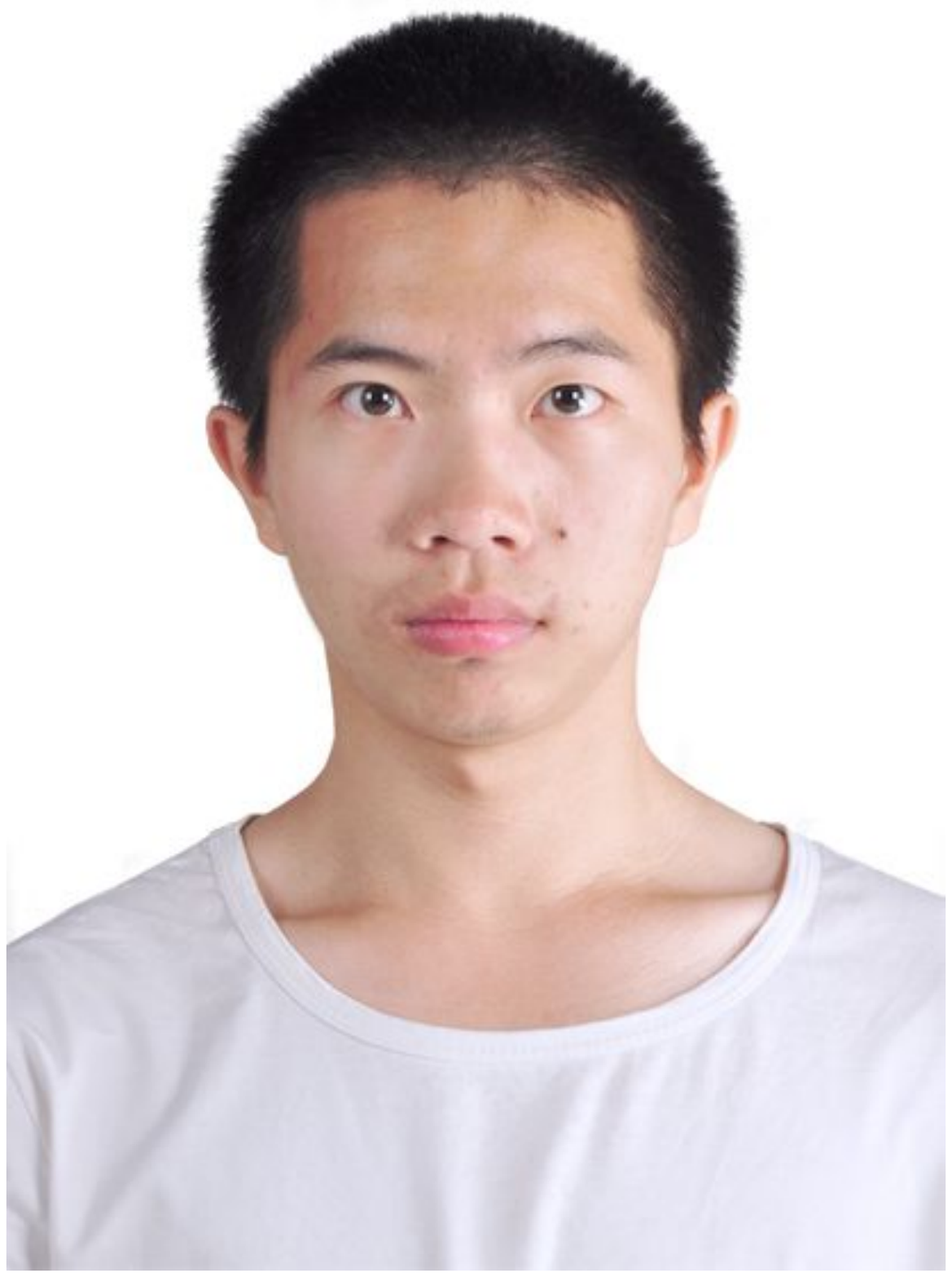}}]{Jia-Hong Xu}
received his B.S. degree in the School of Control and Computer Engineering from North China Electric Power University, China, in 2018. He is currently pursuing his Ph.D. degree in the School of Computer Science and Technology, Huazhong University of Science and Technology, China. His research interests mainly include non-volatile memory and in-memory computing.
\end{IEEEbiography}

\begin{IEEEbiography}[{\includegraphics[width=1in,height=1.25in,clip,keepaspectratio]{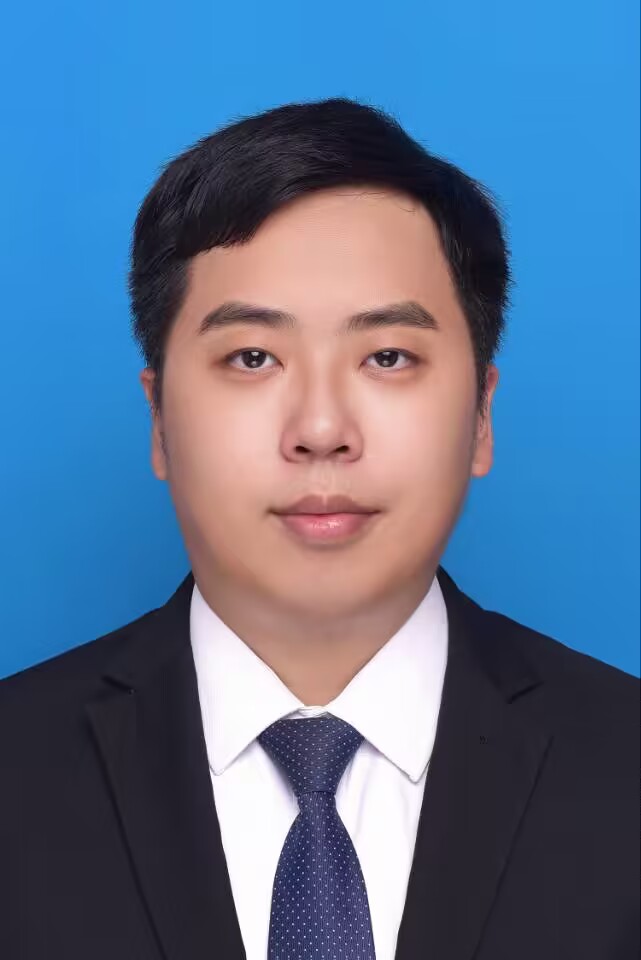}}]{Zhuo-Hui Duan}
is currently a postdoctoral fellow in the School of Computer Science and Technology, HUST, in China. He has presented papers in several conferences such as ASPLOS,SC,DAC,ICDCS,ICCD,DATE. His research interests are in hybrid memory and distribute memory pool.
\end{IEEEbiography}

\begin{IEEEbiography}[{\includegraphics[width=1in,height=1.25in,clip,keepaspectratio]{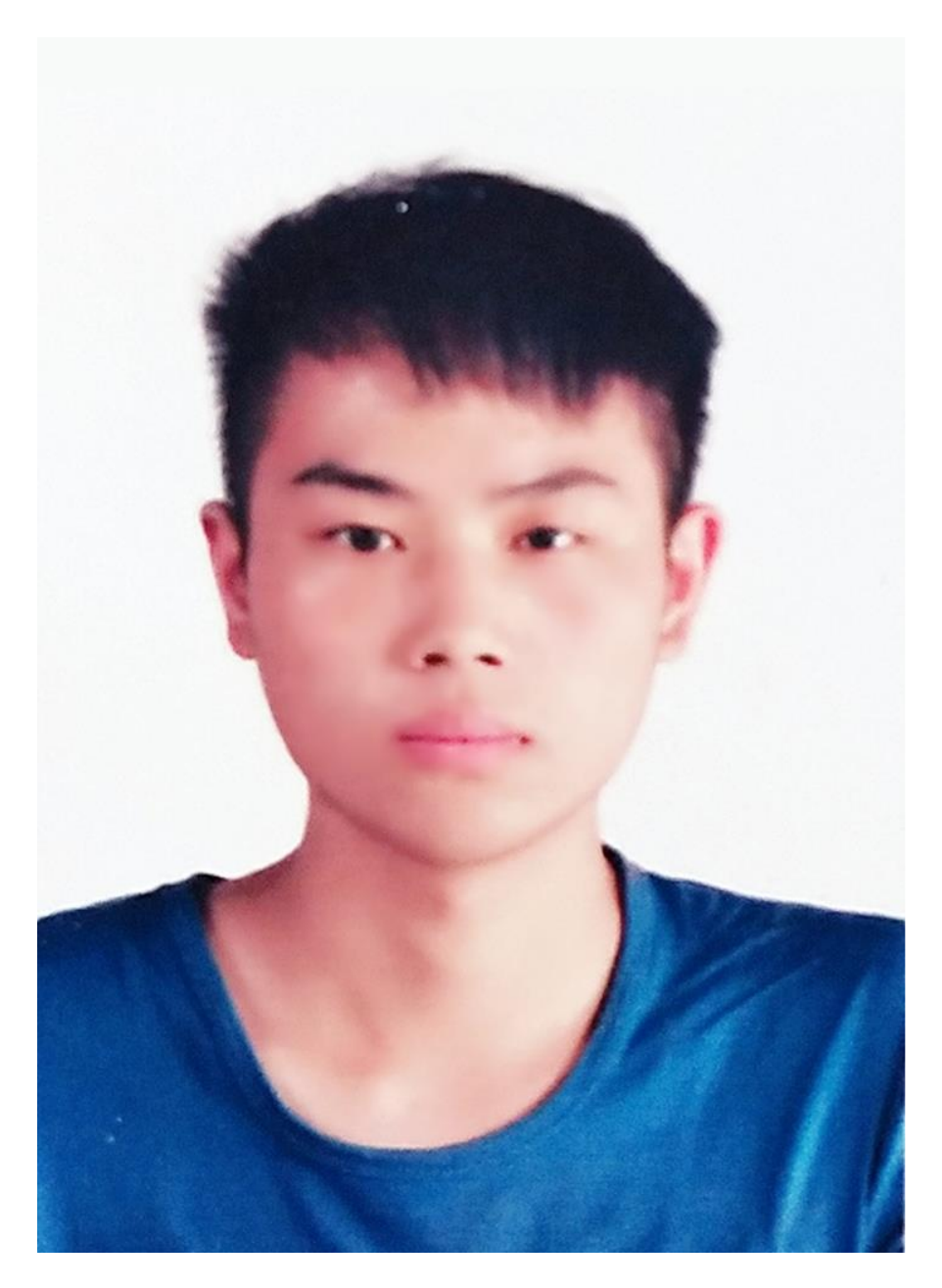}}]{Dan Chen}
 received the BS degree from the
North China Electric Power University, in 2018. He is now working toward the PhD degree in the School of Computer Science
and Technology, HUST, in China. His research
interests focus on processing-in-memory and graph neural network.
\end{IEEEbiography}

\begin{IEEEbiography}[{\includegraphics[width=1in,height=1.25in,clip,keepaspectratio]{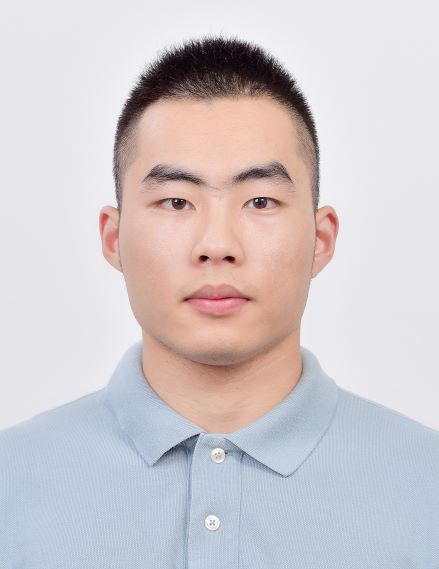}}]{Chuang-Yi Gui}
is currently a Ph.D. candidate in the School of Computer Science and Technology at Huazhong University of Science and Technology (HUST), Wuhan, China. He received his B.E. degree at HUST in 2017. His current research interests include graph processing and graph mining systems.
\end{IEEEbiography}

\end{document}